\documentclass[twocolumn,10pt]{IEEEtran}
\usepackage{comment}
\usepackage{algorithmic}
\usepackage{textcomp}
\includecomment{toexclude}
\usepackage{amsthm}
\usepackage{amsmath,amsfonts,amssymb}
\usepackage{cite}
\usepackage{verbatim}
\usepackage{bm}
\usepackage{graphicx}
\usepackage{xcolor}
\usepackage{subcaption}
\usepackage{epstopdf}
\usepackage{mathrsfs}
\usepackage{txfonts}
\usepackage{lineno}
\usepackage{multirow}
\usepackage{booktabs}
\usepackage{epstopdf}
\usepackage{color}
\usepackage{multicol}
\usepackage{stfloats}
\usepackage{diagbox}
\usepackage{esint}
\usepackage{float}
\usepackage[linesnumbered,ruled,vlined]{algorithm2e}
\def\BibTeX{{\rm B\kern-.05em{\sc i\kern-.025em b}\kern-.08em
		T\kern-.1667em\lower.7ex\hbox{E}\kern-.125emX}}
	
\ifCLASSINFOpdf \else \fi
\graphicspath{{./images/}}

\newcommand{\xt}{\mathbf{x}}

\newcommand{\Ht}{\mathbf{H}}
\newcommand{\hf}{\mathbf{h}}
\newcommand{\Phit}{\mathbf{\Phi}}

\newcommand{\diag}{\text{diag}}

\definecolor{mygreen}{RGB}{32,178,170}  

\definecolor{mygolden}{RGB}{255,140,0} 

\begin{document}
\title{Learning-based Intelligent Surface Configuration, User Selection, Channel Allocation, and Modulation Adaptation for Jamming-resisting Multiuser OFDMA Systems\\}
\author{ Xin Yuan,~\IEEEmembership{Member, IEEE}, Shuyan Hu,~\IEEEmembership{Member, IEEE}, Wei Ni,~\IEEEmembership{Senior Member, IEEE},\\ Ren Ping Liu,~\IEEEmembership{Senior Member, IEEE},
and Xin Wang,~\IEEEmembership{Fellow, IEEE}
	\thanks{
        X. Yuan and W. Ni are with CSIRO, Sydney, Australia.\par
		S. Hu and X. Wang are with Fudan University, Shanghai, China.\par
	    R. P. Liu is with the Global Big Data Technologies Center, University of Technology Sydney, Sydney, Australia.\par	
}}
\maketitle

\begin{abstract} 
Reconfigurable intelligent surfaces (RISs) can potentially combat jamming attacks by diffusing jamming signals. This paper jointly optimizes user selection, channel allocation, modulation-coding, and RIS configuration in a multiuser OFDMA system under a jamming attack. This problem is non-trivial and has never been addressed, because of its mixed-integer programming nature and difficulties in acquiring channel state information (CSI) involving the RIS and jammer. We propose a new deep reinforcement learning (DRL)-based approach, which learns only through changes in the received data rates of the users to reject the jamming signals and maximize the sum rate of the system. The key idea is that we decouple the discrete selection of users, channels, and modulation-coding from the continuous RIS configuration, hence facilitating the RIS configuration with the latest twin delayed deep deterministic policy gradient (TD3) model. Another important aspect is that we show a winner-takes-all strategy is almost surely optimal for selecting the users, channels, and modulation-coding, given a learned RIS configuration. Simulations show that the new approach converges fast to fulfill the benefit of the RIS, due to its substantially small state and action spaces. Without the need of the CSI, the approach is promising and offers practical value. 
\end{abstract}


\IEEEpeerreviewmaketitle

\section{Introduction}\label{sec-intro}
Jamming attacks are severe security threats to wireless systems owing to the broadcast nature of radios~\cite{Zou2016,Amuru2015, yuan2019Secrecy, yuan20}. Many techniques have been adopted to defend against jamming attacks, such as beamforming, frequency hopping, and power control~\cite{Gao2018Game, Hanawal2016, Zhou2017Jamsa}.
Reprogrammable metasurfaces, also known as reconfigurable smart surfaces (RISs), are one of the emerging technologies for wireless systems that have been proposed to combat interference~\cite{Yang2021Intelligent,Renzo2020Smart,Basar2019Wireless}. 
For example, RIS was considered to empower smart radio environments \cite{Renzo2020Smart} and facilitate wireless communications \cite{Basar2019Wireless}.
An RIS is typically composed of densely placed, low-cost, passive meta-atoms, and can reconfigure the radio propagation environments between a transmitter-receiver pair, by fine-tuning the phase shifts of the passive meta-atoms to produce favorable scatterings and reflections~\cite{Renzo2020Smart,Basar2019Wireless,Wu2019Intelligent,Li2021Intelligent}.

\begin{figure}[ht]
	\centering
	\includegraphics[width=0.9\columnwidth]{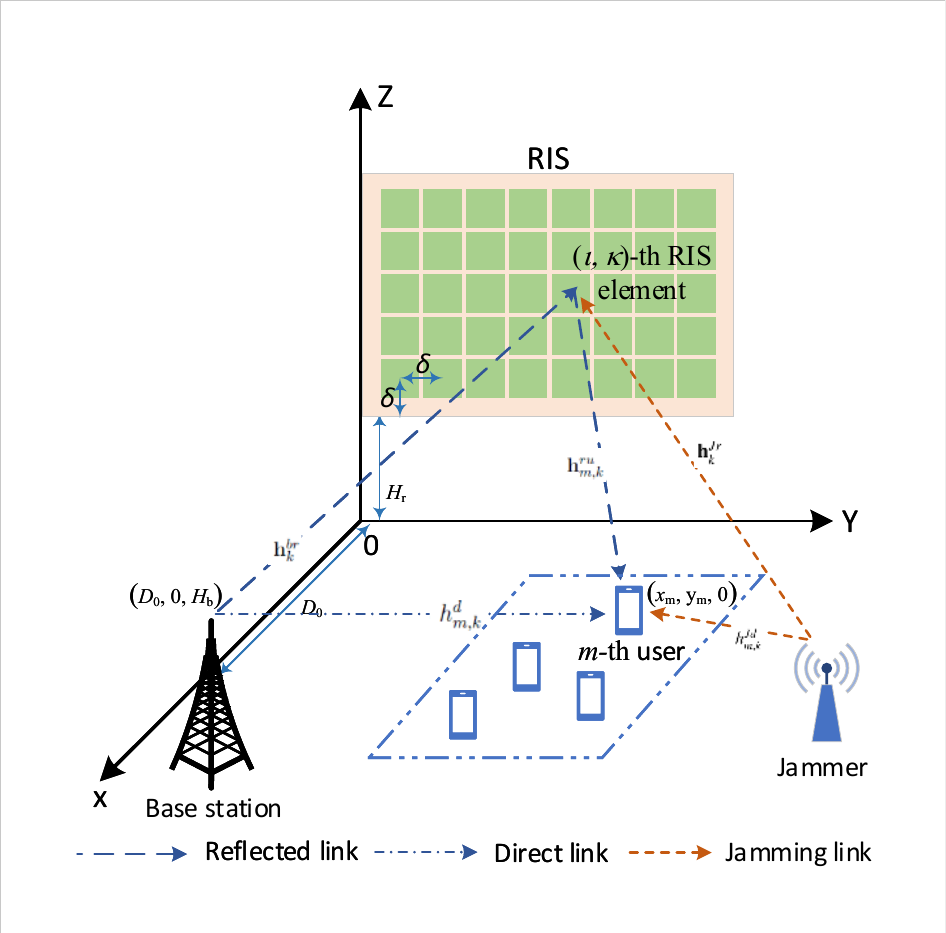}
	\caption{An illustration on an RIS-assisted, downlink multiuser OFDMA system under a jamming attack.}
	\label{fig-sysmodel}
\end{figure}

Fig.~\ref{fig-sysmodel} depicts a generic downlink scenario of a multiuser Orthogonal Frequency Division Multiple Access (OFDMA) system under a jamming attack. 
An RIS is deployed to help the users reject the jamming signals and enhance the desired signals. 
It is crucial to holistically design the user selection, channel allocation, modulation-coding, and RIS configuration by comprehensively considering discrete modulation-coding modes and potentially multiple data streams with diverse quality-of-service (QoS) per user.
It is also critical that the design does not rely on the assumed availability of channel knowledge, especially the channels involving the jammer, as opposed to many existing studies~\cite{Shen2019Secrecy, Chu2020, Cui2019Secure}.  
The user and modulation-coding selection, channel allocation, and RIS configuration are expected to be optimized by exploration and exploitation in the absence of channel knowledge.

The motivation of this paper is to design a practical approach to user scheduling, subchannel assignment, power allocation, and RIS configuration for an emerging RIS-assisted, downlink, multiuser OFDMA system, under prominent practical constraints arising from the difficulty in estimating the channels to and from the RIS, and from the mixed integer programming nature of the problem. 
Considering a generic scenario, we assume that each user can have multiple data streams with different quality requirements (e.g., the base and enhancement layers of video traffic). We also assume that there can be an intentional jamming device (or an unintentional interference source) in the system. 
The problem is new and challenging. 
To the best of our knowledge, the problem has never been studied in the existing literature. 

\subsection{Related Work}
Many existing studies on RIS-assisted, secure wireless systems have assumed that the base station (BS) possesses perfect and instantaneous CSI of individual channels, including those involving the RISs, and jammers or eavesdroppers~\cite{Shen2019Secrecy, Chu2020, Cui2019Secure}.
Typical solvers, such as alternating optimization~(AO), semidefinite relaxation (SDR), fixed-point iteration method, and block-coordinate descent (BCD), have been applied to obtain approximate solutions~\cite{Shen2019Secrecy, Chu2020, Cui2019Secure, Hong2021Artificial}.
AO was used to devise the beamformer of the BS and the phase shifts of the RIS, to optimize the secrecy rate of an RIS-assisted, secure MISO system~\cite{Shen2019Secrecy}. 
SDR was used to optimally configure the RIS and allocate the power of the BS to enhance the secrecy rate in the existence of an eavesdropper~\cite{Chu2020}.
In~\cite{Cui2019Secure}, both AO and SDR were adopted to improve the secrecy rate by optimally assigning the transmit beamformer and configuring the RIS.  
In~\cite{Hong2021Artificial}, BCD was used to optimize the beamformer and artificial noise~(AN) covariance matrix of the BS and the phase shifts of the RIS, thereby maximizing the sum rate of AN-aided multiple-input multiple-output (MIMO) systems.

Considering imperfect CSI, the authors of \cite{Yu2020Robust, Ge2021Robust, Wang2021Distributionally,Sun2021Intelligent}
provided robust designs of the RIS and the BS's beamformers in the presence of jammers or eavesdroppers.
In~\cite{Yu2020Robust}, the imperfect CSI was exploited to optimize the transmit beamformer and AN covariance matrix of the BS, and the phase shifts of the RIS, under the constraint of the maximum allowed information leakage.
In~\cite{Ge2021Robust}, active and passive secure beamforming techniques were developed under a deterministic CSI error model.
In~\cite{Wang2021Distributionally}, a moment-based random error model was used to model CSI errors, followed by optimizing the secure beamformer of the BS and the RIS configuration.
The authors of~\cite{Sun2021Intelligent} maximized the sum rate by jointly designing the BS's transmit beamformer and configuring the RIS without knowing the jammer's transmit beamformer, when there is a jammer and an eavesdropper.
The bounded CSI error model of a third-party node was assumed over each link. The error bounds were known to the BS. 
These methods~\cite{Yu2020Robust, Ge2021Robust, Wang2021Distributionally,Sun2021Intelligent} all needed the statistical CSI.  		

Despite deep reinforcement learning (DRL) has been increasingly applied to wireless communication systems, e.g., spectrum sensing~\cite{Chang2022DEQN}, mobile edge computing~\cite{2020XiaoReinforcement}, and resource allocation~\cite{2022HazarikaDRL}, only a few works have adopted DRL for RIS-assisted secure communications, i.e.,~\cite{Yang2021Intelligent} and~\cite{Yang2021Deep}. The phase shifts of the RIS were also discretized to produce a discrete action space in the few studies~\cite{Yang2021Intelligent, Yang2021Deep}.
Unfortunately, none of these existing studies can apply to the problem at hand, 
due to the complex and mixed integer programming nature of the problem (with the continuous RIS configuration and discrete selection of the user, data stream, subchannel, and modulation-coding mode). 
In~\cite{2020XiaoReinforcement}, a DRL-based mobile offloading scheme was proposed for edge computing against jamming. In the scheme, an actor network chooses continuous offloading policies. A critic network updates the actor network weights to improve the computational performance without knowing the task generation model, edge computing model, and jamming model. Although the continuous action spaces were considered, the problem studied in [R6] did not consider an RIS and is substantially different from this paper.

As found in~\cite{Choi2011Implementation}, adaptive modulations of grouped subcarriers can improve OFDM performance in millimeter wave (mmWave) frequencies.
In \cite{Jian2021Baseband}, waveform and modulation-coding were designed to lower the peak-to-average-power ratio of terahertz transmissions.
In~\cite{morales2021adapt}, modulation-coding was adapted to the received power of terahertz signals.
In \cite{Zhang2020Augmenting}, an adaptive modulation-coding mechanism was developed for a tunable reflector-assisted mmWave system. The outage probability and throughput of the mechanism
were analyzed using stochastic geometry. 
However, these studies~\cite{Choi2011Implementation, Jian2021Baseband, morales2021adapt, Zhang2020Augmenting} were restricted to a single-user setting, and cannot apply to the new multiuser scenario considered in this paper.

\subsection{Contribution and Organization}
In this paper, we jointly optimize the user selection, channel allocation, 
modulation-coding, and RIS configuration for an RIS-assisted downlink multiuser OFDMA system under a jamming attack. 
A new DRL-based approach is developed, which does not require the CSI knowledge of individual links and learns only through the changes in the readily available received data rates of the users to configure the RIS, reject the jamming signals, support diverse data qualities, and maximize the sum rate of the system.
The key contributions of this paper are listed, as follows.
\begin{itemize}
	\item A new problem is considered to comprehensively optimize the user and modulation-coding selection, channel allocation, and RIS configuration in a downlink multiuser OFDMA system under a jamming attack. The RIS is configured to diffuse the jamming signals and direct the desired signals to the intended recipients. 	 
	
	\item
	We decouple the continuous RIS configuration from the discrete user and modulation-coding selection and channel allocation. A new twin delayed deep deterministic policy gradient (TD3) model is designed to adaptively configure the RIS by learning only from the changes in the received data rates of the users, hence eliminating the need of CSI knowledge. 	
	\item
	A winner-takes-all strategy is designed to deliver the almost surely optimal user and modulation-coding selection, and channel allocation, hence reducing the action space and contributing to the fast and reliable convergence of the TD3 model. 
\end{itemize}
Extensive simulations confirm that the proposed TD3-based framework significantly outperforms its non-learning alternatives in terms of sum rate.
The gain of a meticulously configured RIS is demonstrated, as the system with 40, 60, or 80 reflecting elements at the RIS provides 16.50\%, 32.91\%, or 51.86\% higher sum rates than the system without the RIS, respectively.
Eliminating the need of CSI, the proposed framework is of significant practical value.

The remainder of this paper is arranged as follows.
Section \ref{sec-model} sets forth the system model. 
Section \ref{sec-ddpg} articulates with the new TD3-based framework for joint user and modulation-coding selection, channel allocation, and RIS configuration.
In Section~\ref{sec-sim}, the new framework is numerically evaluated,
followed by conclusions in Section~\ref{sec-con}.
Notations used in the rest of the paper are collated in Table~\ref{table_notations}.
\begin{table}[t]
	\caption{Notation and Definition}
	\renewcommand\tabcolsep{1.2pt}
	\begin{center}
		\begin{tabular}{ll}
			\toprule[1.5pt]
			Notation & Definition \\
			\hline
			$M$ & Number of users \\
			$\cal M$ & Set of users \\
			$N$ & Number of the reflecting elements of the RIS\\
			${\cal N}$ & Set of the reflecting elements of the RIS\\
			$K$ & Number of subchannels\\
			${\cal K}$ & Set of subchannels\\
			$L$ & Number of modulation-coding modes\\
			${\cal L}$ & Set of modulation-coding modes\\
			$Q$ & Number of data streams\\
			${\cal Q}$ & Set of data streams\\
			$\xt_k$ & Transmit symbols in the $k$-th subchannel\\
			$h^{d}_{m,k}$ & Channel coefficient from BS to the $m$-th user in the $k$-th \\ &  subchannel\\
			$\hf^{ru}_{m,k}$ & Channel vector from RIS to the $m$-th user in the $k$-th\\ & subchannel\\
			$\hf^{br}_k$ & Channel matrix from BS to RIS in the $k$-th subchannel\\
			$h^{br}_{k} (n)$ & Channel coefficient from BS to the $n$-th \\ &reflecting element of the RIS\\
			$\hf^{Jr}_k$ & Channel matrix from jammer to RIS in the $k$-th \\ &  subchannel\\
			$h^{Jr}_{k} (n)$ & Channel coefficient from jammer to the $n$-th reflecting\\ & element of the RIS\\
			$\Phit$ & The RIS's reflection matrix $\Phit \triangleq \diag\{\phi_{1},\cdots,\phi_{N}\}$\\
			$\phi_{n}$ & Reflection coefficient of the $n$-th reflecting element \\ & of the RIS \\
			$\theta_{n}$ & Phase shift of the $n$-th reflecting element of the RIS\\
			$h^{Jd}_{m,k}$ & Channel coefficient from the jammer to the $m$-th user\\ & in the $k$-th subchannel\\
			$h_{m,k}$ & Effective channel from the BS to the $m$-th user in \\ & the $k$-th subchannel\\
			$h^J_{m,k}$ & Effective channel from the jammer to the $m$-th user \\ & in the $k$-th subchannel\\
			${n}_{m,k}$ & CSCG noise with zero mean and variance $\sigma^2$\\
			$\eta^{(q)}_{m,k,l}\left(|h_{m,k}|^2\right)$ & Indicator for the selection of the $k$-th subchannel and \\ & the $l$-th modulation-coding mode to deliver the $q$-th\\ &  data stream of the $m$-th user \\
			$r_l$ & Transmit rate of the $l$-th modulation-coding mode\\
			$p^{(q)}_{m,k,l}\left(|h_{m,k}|^2\right)$ & Minimum transmit power required for the BS to deliver\\ &  the $q$-th data stream of the $m$-th user in $k$-th subchannel\\ & the  using the $l$-th modulation-coding mode \\
			$P_{\max}$ & Maximum transmit power of the BS\\
			$P_J$ & Transmit power of the jammer \\
			$\varrho_{m,k,l}$ & Bit error rate (BER) of the $m$-th user in the $k$-th \\ & subchannel using the $l$-th modulation-coding mode\\
			$\varrho_0$ & BER requirement for all data streams of all users\\
			\toprule[1.5pt]
		\end{tabular}
	\end{center}
	\label{table_notations}
\end{table}

\section{System Model}\label{sec-model}
We study an RIS-assisted downlink multiuser OFDMA system, where a single-antenna BS serves $M$ single-antenna users via $K$ orthogonal subchannels, as illustrated in Fig.~\ref{fig-sysmodel}.
A malicious single-antenna jammer is located near the users and
sends jamming signals in an attempt
to block the legitimate receptions of the users.
An RIS comprising a uniform rectangle array (URA) of $N = N_y \times N_z$ reflecting elements is installed on the facade of a building, which is controlled by the BS to help reject/diffuse the jamming signals and enhance the legitimate communications. 
Here, $N_y$ and $N_z$ are the numbers of reflecting elements in each row and column of the RIS, respectively. 
The phase shifts of the RIS's reflecting elements are individually adjustable with a smart controller. 
Denote by ${\cal M} = \{1,\cdots,M\}$, ${\cal K} = \{1,\cdots,K\}$, and ${\cal N} = \{1,\cdots,N\}$ the sets of users, subchannels, and RIS's reflecting elements, respectively.

We consider that the BS configures the RIS and sends pilot signals at the beginning of every block. The users estimate their effective channels, and feed back their achievable rates to the BS. 
The BS selects users, and allocates subchannels and modulation-coding modes for delivery of different data streams to the users in the rest of the block, only based on the users' feedback of their achievable data rates. No explicit CSI of the channels is needed. 
This consideration is practically interesting, due to the difficulty and significant overhead needed to estimate the individual channels involving the RIS or the jammer~\cite{Hu2021Two}. In contrast, the received data rates can be readily measured and reported by the users~\cite{Lin2020Deep}. 

Let ${\xt}_k \triangleq \left[x_{1,k},\cdots,x_{M,k}\right]^T \in {\mathbb{C}^{M \times 1}}$ denote the transmit symbols for the $M$ users in the $k$-th subchannel, and $\xt_k^J \triangleq \left[x^J_{1},\cdots,x^J_{K} \right]^T \in {\mathbb{C}^{K \times 1}} $ denote the jamming signals on the $K$ subchannels.
The jamming signals follow the zero-mean circularly symmetric complex Gaussian (CSCG)
distribution with variance $P_J$ \cite{Amuru2015}.
The received signal at the $m$-th user in the $k$-th subchannel is
\begin{equation}\label{eq-ym}
	\begin{aligned}
	y_{m,k} &= \left[\left(\hf^{ru}_{m,k} \right)^H \Phit \hf^{br}_k + h^d_{m,k} \right] \sqrt{p_{m,k}} x_{m,k} \\
	&\quad+ \left[\left(\hf^{ru}_{m,k} \right)^H \Phit \hf^{Jr}_k + h^{Jd}_{m,k} \right] \sqrt{p_{k}^{J}} x^{J}_{k} + n_{m,k}, \forall m,k,
	\end{aligned}	
\end{equation}
where $h^{d}_{m,k}$ is the channel coefficient from BS to the $m$-th user in the $k$-th subchannel; $\hf^{ru}_{m,k} = \left[ h^{ru}_{m,k} (1),\cdots, h^{ru}_{m,k} (N) \right]^T \in \mathbb{C}^{N \times 1}$ is the channel vector from RIS to the $m$-th user in the $k$-th subchannel;
$\hf^{br}_k= \left[h^{br}_{k} (1),\cdots, h^{br}_{k} (N)\right]^T \in \mathbb{C}^{N \times 1}$ is the channel matrix from BS to RIS in the $k$-th subchannel, and $h^{br}_{k} (n), \;n \in {\cal N}$ is the channel coefficient from BS to the $n$-th RIS's reflecting element;
$p_{m,k}$ is the BS's transmit power for the $m$-th user in the $k$-th subchannel;
$h^{Jd}_{m,k}$ is the channel coefficient from the jammer to the $m$-th user in the $k$-th subchannel;
$\hf^{Jr}_k = \left[h^{Jr}_{k} (1),\cdots, h^{Jr}_{k} (N)\right]^T \in \mathbb{C}^{N \times 1}$ is the channel matrix from the jammer to RIS in the $k$-th subchannel, with $h^{Jr}_{k} (n), \;n \in {\cal N}$ being the channel coefficient from the jammer to the $n$-th RIS's reflecting element;
$p^J_{k}$ is the transmit power of the jammer in the $k$-th subchannel;
${n}_{m,k} \in {\cal CN}\left(0,\sigma^2 \right), \forall m \in {\cal M},\;k \in {\cal K}$ is the zero-mean CSCG noise with variance $\sigma^2$; 
and $\Phit \triangleq \diag\{\phi_{1},\cdots,\phi_{N}\}$ is the RIS's reflection matrix. 
$\phi_{n} = e^{j\theta_{n}}$ is the reflection coefficient of the $n$-th reflecting element of the RIS with
$\theta_{n} \in [0,2\pi)$ being the phase shift of the reflecting element, and $\left|\phi_{n} \right| \leq 1$.
Being a transmitting device, the jammer can be hardly aware of the user selection at each subchannel. It is reasonable for the jammer to transmit its full power across the spectrum to block the users.

The effective channel coefficients from the BS or jammer to the $m$-th user in the $k$-th subchannel are given by 
\begin{equation}
	h_{m,k} = \left(\hf^{br}_k \right)^H \Phit^H \hf^{ru}_{m,k} + h^d_{m,k}, \; \forall m \in {\cal M},\; k \in {\cal K};
\end{equation}
\begin{equation}
h^J_{m,k} = \left(\hf^{Jr}_k \right)^H \Phit^H \hf^{ru}_{m,k} + h^{Jd}_{m,k}, \; \forall m \in {\cal M},\; k \in {\cal K}.
\end{equation}
Suppose that the channels undergo block fading, i.e., the channels are unchanged within a block and vary independently between blocks~\cite{Viswanathan1999Capacity}. 
The received signal-to-interference-plus-noise ratio (SINR) at the $m$-th user in the $k$-th subchannel is 
\begin{equation}
	\gamma_{m,k} = \frac{ p_{m,k}|h_{m,k}|^2}{p^J_{k}|h^J_{m,k}|^2 + \sigma^2}.
\end{equation}

The BS can select the $l$-th modulation-coding mode from a discrete set of modulation-coding modes ${\cal L}$, and the corresponding transmit rate is $r_l, \;l \in {\cal L} = \{0, 1, \cdots, L\}$. The number of available modulation-coding modes is $L=|\mathcal{L}|$, where $|\cdot|$ stands for cardinality.
Note that $l = 0$ indicates no transmission, i.e., $r_0 =0$.
By employing the $l$-th modulation-coding mode, the BER at the $m$-th user in the $k$-th subchannel is~\cite{Goldsmith1998}
\begin{equation}\label{eq-ber}
	\varrho_{m,k,l} = \beta_1 \exp\left(\frac{\beta_2 \gamma_{m,k}}{2^{r_l} - 1} \right), 
\end{equation}
where $\beta_1$ and $\beta_2$ are constants depending on the modulation-coding scheme.

We also consider $Q$ data streams with different BER requirements for each user. The index to the data streams is $q\in {\cal Q} = \{1,\cdots,Q\}$. For illustration convenience, we set $Q =2$, where $q=1$ indicates high-quality (HQ) data streams and $q=2$ indicates low-quality~(LQ) data streams.
To meet the BER requirements $\varrho^{(q)}_0, q \in {\cal Q}$ (or in other words, the QoS requirements) of the $q$-th data stream, the minimum transmit power required for the BS to deliver the $q$-th data stream of the $m$-th user in the $k$-th subchannel using the $l$-th modulation-coding mode is~\cite{Malik2018Interference}
\begin{equation}\label{eq-power}
p^{(q)}_{m,k,l}\left(|h_{m,k}|^2, |h^J_{m,k}|^2\right) = \frac{\left(2^{r_l} -1 \right) \ln\left(\frac{\beta_1}{\varrho^{(q)}_0} \right)\left(p^J_{k}|h^J_{m,k}|^2 + \sigma^2 \right) }{\beta_2 \left|h_{m,k}\right|^2}.
\end{equation}

\section{Proposed Channel Allocation, Modulation-coding selection, and RIS Configuration}\label{sec-ddpg}
The BS assigns the subchannels for the users, select the modulation-coding modes, and configures the RIS.
Let $\eta^{(q)}_{m,k,l}(|h_{m,k}|^2, |h^J_{m,k}|^2 ) = 1$ indicate the selection of the $k$-th subchannel and the $l$-th modulation-coding mode for transmitting the $q$-th data stream of the $m$-th user, given $|h_{m,k}|^2$ and $|h^J_{m,k}|^2$; and $\eta^{(q)}_{m,k,l} = 0$ indicates otherwise.
Let $\bm{\eta} := \left\lbrace \eta^{(q)}_{m,k,l}(|h_{m,k}|^2, |h^J_{m,k}|^2), \;\forall m \in {\cal M},\;k \in {\cal K},\;l \in {\cal L},\;q \in {\cal Q}\right\rbrace $ collect all indicators.
The transmit rate for delivering the $q$-th data stream of the $m$-th user in the $k$-th subchannel is
\begin{equation}
	R^{(q)}_{m,k}\left(\bm{\eta}\right) =  \sum_{l=0}^{L} \eta^{(q)}_{m,k,l}\left(|h_{m,k}|^2,|h^J_{m,k}|^2 \right) r_l.
\end{equation}
The sum rate of the system is given by
\begin{equation}\label{eq-sum_rate}
\begin{aligned}
     R_{\rm{tot}}\left(\bm{\eta}\right) = \!\sum_{m = 1}^{M}\! R_m \left(\bm{\eta}\right)
     = \!\sum_{m = 1}^{M} \sum_{k = 1}^{K}\sum_{l=0}^{L}\sum_{q=1}^{Q}\! \eta^{(q)}_{m,k,l}\left(|h_{m,k}|^2, |h^J_{m,k}|^2 \right) r_l,
\end{aligned}	
\end{equation}
where $R_m \left(\bm{\eta}\right) = \sum_{k = 1}^{K}\sum_{l=0}^{L}\sum_{q=1}^{Q} \eta^{(q)}_{m,k,l}\left(|h_{m,k}|^2, |h^J_{m,k}|^2 \right) r_l$ is the data rate received at the $m$-th user.

The total transmit power of the BS for the $m$-th user is
\begin{equation}
P_m\left({\bm{\eta}}\right) = \!\sum_{k = 1}^K \sum_{l=0}^{L}\sum_{q=1}^{Q}\! \eta^{(q)}_{m,k,l}\left(|h_{m,k}|^2, |h^J_{m,k}|^2 \right) p^{(q)}_{m,k,l}\left(|h_{m,k}|^2, |h^J_{m,k}|^2 \right),\forall m. 
\end{equation}
The total transmit power of the BS is 
\begin{equation}
	P\left({\bm{\eta}}\right) = \!\sum_{m = 1}^M \sum_{k = 1}^K \sum_{l=0}^{L}\sum_{q=1}^{Q}\! \eta^{(q)}_{m,k,l}\left(|h_{m,k}|^2, |h^J_{m,k}|^2 \right) p^{(q)}_{m,k,l}\left(|h_{m,k}|^2, |h^J_{m,k}|^2 \right). 
\end{equation}

We jointly design the selection of channels, user and modulation-coding modes, $\bm{\eta}$, and the configuration of the reflection matrix of the RIS, $\Phit \in \mathbb{C}^{N \times N}$, to maximize the sum rate of the system while meeting the BER requirements of the users, $\varrho_0^{(q)},\,\forall q\in \mathcal{Q}$. The transmit power of the BS is upper bounded by $P_{\max}$. 
The problem is cast as 
\begin{subequations}\label{eq-P1}
	\begin{align}
	\textbf{P1}:\;\max_{\{\Phit, \bm{\eta} \}}\;\; & R_{\rm{tot}}\left({\bm{\eta}} \right) \\ 
	{\text{s.t.}}\; & P\left({\bm{\eta}}\right) \leq P_{\max},\label{eq-P1 b}\\
	& \theta_{n} \in [0,2\pi), \forall n \in {\cal N}, \label{eq-P1 c}\\
	&\sum_{k=1}^{K}\left\lbrace \sum_{m = 1}^M\sum_{l=0}^{L}\sum_{q=1}^{Q} \eta^{(q)}_{m,k,l} \left(|h_{m,k}|^2, |h^J_{m,k}|^2\right)\right\rbrace  \leq K,\label{eq-P1 d}\\
	&\sum_{m = 1}^M\sum_{l=0}^{L}\sum_{q=1}^{Q} \eta^{(q)}_{m,k,l} \left(|h_{m,k}|^2, |h^J_{m,k}|^2\right) \leq 1,\label{eq-P1 e}\\
	& \eta^{(q)}_{m,k,l} \left(|h_{m,k}|^2, |h^J_{m,k}|^2\right) \in \{0,1\}, \label{eq-P1 f}\\
	& R^{(1)}_m\left(\bm{\eta}\right) = \chi R^{(2)}_m\left(\bm{\eta}\right). \label{eq-P1 g}
	\end{align}
\end{subequations}
Constraint~\eqref{eq-P1 c} specifies the range of the RIS's phase shifts; 
\eqref{eq-P1 d} indicates the number of subchannels assigned to all users is no larger than~$K$;
\eqref{eq-P1 e} indicates that each subchannel is assigned to no more than one user to prevent inter-user interference. Once $\eta^{(q)}_{m,k,l}$ is determined, the transmit power for the $m$-th user in the $k$-th subchannel using the $l$-th modulation-coding mode, i.e., \eqref{eq-power}, is specified to meet the BER requirement. 
In~\eqref{eq-P1 g}, $\chi$ is the ratio of the HQ and LQ data streams, which needs to be maintained between the streams, e.g., for streaming videos with layered coding~\cite{He2014Optimal}.
\begin{figure*}[tb]
	\centering
	\includegraphics[width=1.6\columnwidth]{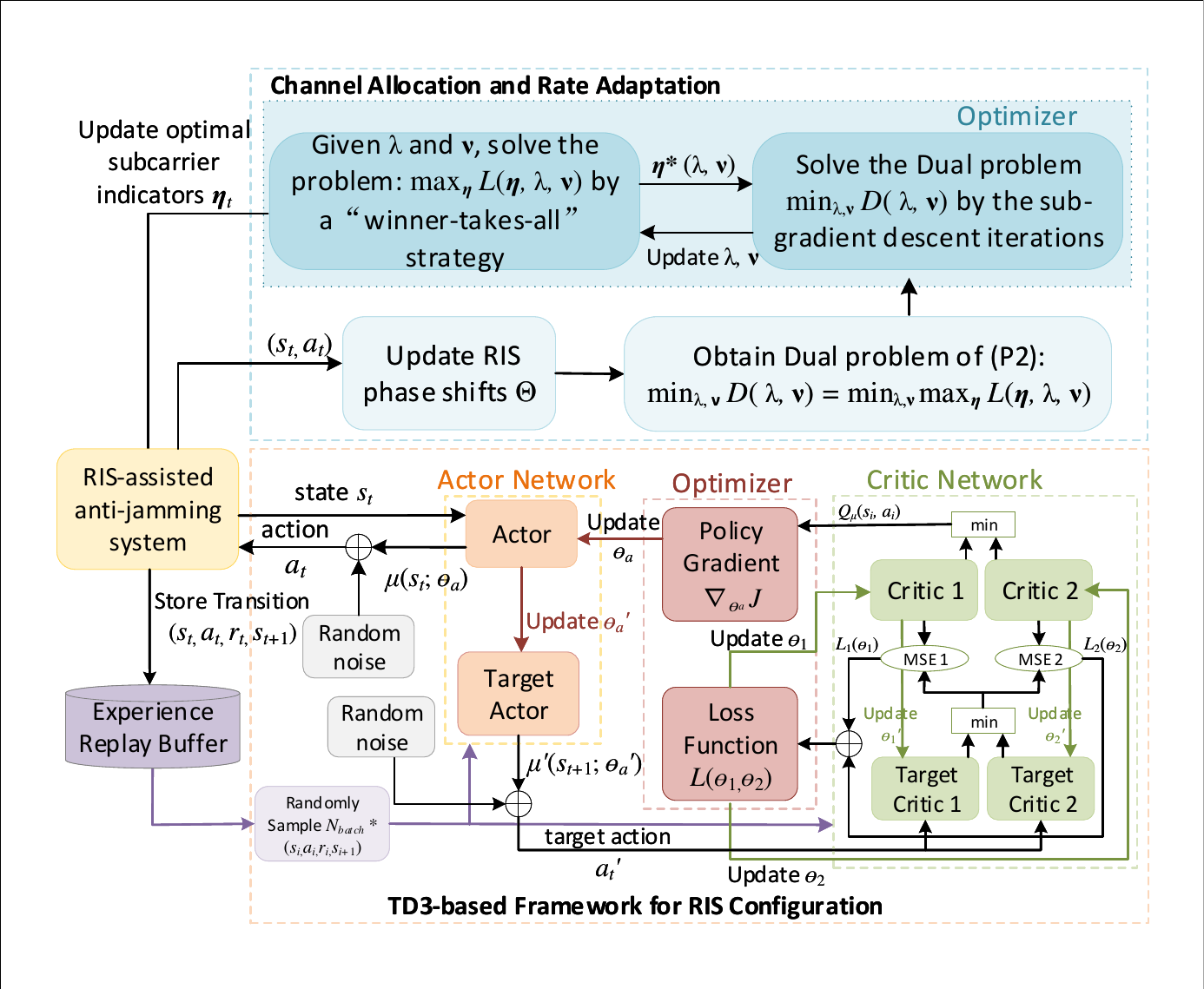}
	\caption{An overview of the proposed TD3-based framework for jointly optimizing the user selection, channel allocation, modulation-coding, and RIS configuration. The top of the figure optimizes the discrete user and modulation-coding selection and channel allocation using primal-dual subgradient descent, given the RIS configuration. The bottom optimizes the RIS configuration using TD3, given the outcome of the top. }
	\label{fig-algorithm}
\end{figure*}

Problem \textbf{P1} is a non-convex combinatorial problem, and intractable for conventional optimization techniques.
We design a new framework to solve the problem, which configures the RIS by using DRL. Given a possible configuration of the RIS, we rely on primal-dual subgradient descent (PSD) to optimize the allocations of subchannels and modulation-coding modes by using a winner-takes-all strategy. 
By iteratively configuring the RIS and optimizing the allocations, the framework can substantially reduce the state and action spaces of the DRL and quickly converge to a superb solution. 

\subsection{Twin Delayed DDPG (TD3)-based RIS Configuration}\label{subsec-ddpg}
Problem \textbf{P1} can be decoupled into sequential decisions of the RIS configuration, the channel allocation, and the user and modulation-coding selection. Specifically, given an RIS configuration, the received data rates of the users are readily available. The channel allocation, user and modulation-coding selection only depend on the received data rates. The RIS configuration involves $N$ constant-modulus variables, i.e., $\theta_n,\,\forall n\in \mathcal{N}$. The extensively adopted solver, SDR, requires the CSI knowledge of all channels, including those involving the RIS; and can only configure the RIS approximately and suboptimally, due to the need of rank randomization~\cite{Wang2020Robust}.

The benefit of this decoupled learning and optimization structure of the proposed algorithm is two-fold. 
\begin{itemize}
    \item On the one hand, the need for the instantaneous CSI to and from the IRS is eliminated. The users only need to estimate their effective end-to-end channels based on the pilot signals of the BS, e.g., by using the minimum mean square estimation (MMSE), as done in typical wireless communication systems, e.g., 3GPP LTE. By this means, we can circumvent the impasse of estimating the CSI to and from the RIS.  
    
    \item On the other hand, the primary subgradient descent-based selections of the user, data stream, subchannel, and modulation-coding mode, and power allocation evaluate precisely the maximum reward that can be offered by a given IRS configuration. Under the given RIS configuration, the optimality of the selections is proved rigorously by showing that the selections follow an almost surely unique and optimal ``winner-takes-all'' strategy; see Section \ref{subsec-optimization}. Not only do the optimal selections reduce the state and action spaces of the DRL (which is conducive to the convergence and reliability of the DRL), but ensure the quality of the solution produced by our approach. 
    
\end{itemize}

DRL is an effective dynamic programming tool to solve a sequential decision-making problem by learning optimal solutions in a dynamic environment. In this paper, we employ the DRL to configure the RIS. 
Let the BS serve as the agent. The key elements of the DRL model are specified below. 

{\textbf{\emph{State Space ${\cal S}$: }}}At the $t$-th learning step, the system state $s_t \in {\cal S}$ is defined as
\begin{equation}\label{eq-state space}
	 s_t = \left\lbrace R_m,\forall m \in {\cal M}  \right\rbrace. 
\end{equation}

{\textbf{\emph{Action Space $\cal A$: }}} The action space collects all possible actions, i.e., $\mathcal A:=\{a_t, \forall t =1, \cdots, N\}$. At the $t$-th learning step, action $a_t$ includes the reflecting coefficients $\{\theta_{n}\}_{n \in {\cal N}}$, i.e., 
\begin{equation}\label{eq-action}
	a_t = \left\lbrace \theta^{(t)}_{n} \in [0, 2\pi), \forall n \in {\cal N}\right\rbrace.
\end{equation}

{\textbf{\emph{Transition probability: }}}Under action $a_t$, the transition probability from state $s$ to state $s'$ is given by 
\begin{equation}\label{eq-transition probability}
P_{a_t}\left(s, s'\right) = \Pr \left( s_{t+1}=s' | s_{t}=s,a_t = a\right).
\end{equation} 

{\textbf{\emph{Policy: }}}The mapping from the state space, ${\cal S}$, to the action
space, ${\cal A}$, is known as a policy, $\pi: {\cal S} \to {\cal A}$, which is a distribution 
${\pi}(a | s) = \Pr\left(a_t = a | s_t =s \right)$ 
over state $s \in~{\cal S}$.

{\textbf{\emph{Reward: }}}The reward function provides positive rewards
at each learning step, denoted by $r_t$, for executing action $a_t$, and is defined as 
\begin{equation}\label{eq-reward}
	r_t = \sum_{m \in {\cal M}} R_{m}\left({\bm{\eta}}\right),
\end{equation}
where $R_m = \sum_{k \in {\cal K}} R_{m,k}$ is the total transmit rate for the $m$-th user.
With a discount coefficient $\gamma\in (0,1)$, the cumulative discounted reward is given by
\begin{equation}\label{eq-reward-dis}
G_t = \sum_{t=0}^{\infty} \gamma^t r_t.
\end{equation}

{\textbf{\emph{Experience: }}}The history experience is defined as $e_t = \left( s_t,a_t,r_t,s_{t+1} \right) $, and memorized in an experience replay buffer, denoted by ${\bm{R}}$. 

The agent perceives the current system state $s_t$, picks an available action $a_t$,
obtains a reward $r_t$, and transits to a new state $s_{t+1}$.
A policy, $a_t = \pi(s_t)$, projects the state $s_t$ to a feasible action.
The agent selects the policy maximizing the accumulated reward $G_t$. 
Given state $s_t$, action $a_t$, and reward $r_t$, an action-value function, i.e., Q-function, is exploited to evaluate $G_t$, as 
$Q_{\pi}(s_t, a_t)=\mathbb E_{\pi}[G_t | s_t, a_t]$.
It satisfies the Bellman Expectation Equation:
\begin{equation}\label{eq-bellman}
\begin{aligned}
Q_{\pi}\left(s_t,a_t \right) & = \mathbb{E}_{r_t, s_{t+1} \sim {\rm\cal E} }\left[r_t + \gamma \mathbb{E}_{a_{t+1}\sim \pi} \left[Q_{\pi}\left(s_{t+1},a_{t+1} \right) \right] \right], 
\end{aligned}	
\end{equation}
where ${\cal E}$ denotes the environment that the agent interacts with.
It is difficult to apply RL directly to obtain the Q-value, $Q(s_t, a_t)$, owing to the continuous state and action spaces.

Being one of the latest DRL models, TD3 is designed for continuous state and action spaces. To address the Q-value overestimation issue of the deep deterministic policy gradient~(DDPG) algorithm,
TD3 introduces \emph{three improvements} over DDPG, i.e., \emph{clipped double-Q learning with two critics}, \emph{target policy smoothing}, and
\emph{delayed policy update}~\cite{dankwa2019twin}.
\begin{itemize}
	\item {\emph{Clipped double-Q learning with two critics: }}TD3 has two critics (i.e., to produce two Q-values), and admits the smaller of the two Q-values to evaluate the target Q-values in the Bellman error loss functions.
	
	\item {\emph{Target policy smoothing: }}TD3 adds noises to the target action and smooths the Q-function value of the actions to make the policy less likely to exploit the errors in the Q-function.
	
	\item {\emph{``Delayed'' policy updates: }}The actors (i.e., policies) are updated less frequently than the critics.
	For example, it was recommended in \cite{dankwa2019twin} that the actors are updated after the critics are updated twice.
\end{itemize}

The TD3-based framework is made up of an actor network and a critic network, where the actor network comprises an actor and a target-actor, and the critic network comprises two critics and two target-critics, as shown in Fig.~\ref{fig-algorithm}. 
Six deep neural network (DNN) approximators are used in the TD3-based network.
The actor with parameters $\theta_a$, denoted by $\mu\left(s_t; \theta_a \right)$, approximates the policy function of the agent and produces the actions.
The two critics with parameters $\theta_1$ and $\theta_2$, denoted by $Q_{1}(s_t,a_t;\theta_1)$ and $Q_{2}(s_t,a_t;\theta_2)$, estimate two action-value functions of the actions produced by the actor, and output the smaller as the action-value function of the actions~\cite{Silver2014ddpg}.
The target-actor with parameter $\theta'_a$, denoted by $\mu'\left(s_t; \theta'_a \right)$, produces the target policy. 
The two target-critics with parameters $\theta'_1$ and $\theta'_2$, denoted by $Q'_{1}(s_t,a_t;\theta'_1)$ and $Q'_{2}(s_t,a_t;\theta'_2)$, 
generate two Q-values, of which the smaller is taken as the target Q-value.

Based on the actor-critic setting, the TD3 network follows the deterministic policy gradient (DPG) theorem~\cite{Silver2014ddpg}
to update the parameters, $\theta_a$, $\theta_1$, $\theta_2$, $\theta'_a$, $\theta'_1$ and $\theta'_2$, and optimize the actions.
The use of the target network (comprising a target-actor and two target-critics) prevents unstable learning arising from using only an actor-critic network (with a single actor and critic)~\cite{Hou2017}.

The BS (i.e., the agent) takes the received data rates of
the users as the current state $s_t$, and passes it to the actor.
Following the DPG theorem~\cite{Silver2014ddpg}, the actor produces the current strategy
by deterministically mapping a state to an action.
The actor approximates the policy function of the agent and chooses an action $a_t$.
A random exploration noise is appended to the action to poise the exploration of new actions and the exploitation of known actions.
The output action is
\begin{equation}
a_t = {\text{clip}}\Big(\mu\left(s_t; \theta_a \right) + \epsilon, a_{\min}, a_{\max} \Big),
\end{equation}
where the noise $\epsilon$ is randomly sampled from a zero-mean Gaussian distribution (GN) 
with variance $\sigma_e^2$, i.e., $\epsilon \sim {\cal N}(0,\sigma_e^2)$;
$\text{clip}(\cdot)$ is a clipping function to limit the actions within $\left[a_{\min}, a_{\max} \right]$
with $a_{\max}$ and $a_{\min}$ being the upper and lower bounds of the actions, respectively.
As the result of action~$a_t$, the agent is rewarded with~$r_t$ and transits to the state $s_{t+1}$. 
The agent perceives the state $s_{t+1}$ and reserves the transition $(s_t, a_t, r_t, s_{t+1})$ in its experience replay buffer ${\bm R}$. 

With the input $(s_t, a_t)$, the two critics evaluate the action-value functions of the selected action~$a_t$, i.e., $Q_{1}\left(s_t, a_t;\theta_1 \right)$ and $Q_{2}\left(s_t, a_t;\theta_2 \right)$. 
By randomly drawing a sampled transition $(s_i, a_i, r_i,s_{i+1})$ from the experience replay buffer $\bm R$, the action-value functions produced by the two critics are approximated by $Q_{1}\left(s_i, a_i;\theta_1 \right)$ and $Q_{2}\left(s_i, a_i;\theta_2 \right)$. 
The lesser of these two approximate action-value functions is chosen as the Q-value of the next state, i.e.,
$Q_{\mu}(s_i, a_i) = \min \left\{Q_{1}\left(s_i, a_i;\theta_1 \right), Q_{2}\left(s_i, a_i;\theta_2 \right) \right\} $.

Given the probability distribution of the parameter $\theta_a$, i.e., $J(\theta_a)$,
the actor network is updated towards the direction in which the estimation improves the strategy fastest.
In other words, $\theta_a$ is updated towards the direction specified by the gradient of $J(\theta_a)$,
which is given by~\cite{Silver2014ddpg}
\begin{subequations}\label{eq-gradient}
	\begin{align}
	\nabla_{\theta_a} J(\theta_a) & = \mathbb{E}_{s \sim \rho^{\mu}} \left[\nabla_{\theta_a} {{Q}}_{\mu}(s_t,\mu(s_t;\theta_a);\theta_k) \right]\label{eq-gradient a}\\
	& =  \mathbb{E}_{s \sim \rho^{\mu}} \left[\nabla_{\theta_a} \mu(s_t;\theta_a) \nabla_{a} {{Q}}_{\mu}(s_t,\mu(s_t;\theta_a);\theta_k) \right],\label{eq-gradient b}
	\end{align}
\end{subequations}
where $ k= 1$ or $2$; \eqref{eq-gradient b} is derived from the chain rule; 
$\rho^{\mu}$ is a discounted state distribution of policy $\mu(s_t;\theta_a)$~\cite{Timothy2016Continuous};
$\nabla_{\theta_a} \mu(s)$ is the gradient of the actor $\mu(s)$ with respect to (w.r.t.) the parameter $\theta_a$; and $\nabla_{a} {{Q}}_{\mu}(s_t,a;\theta_a)$ is the gradient of ${{Q}}_{\mu}(s_t,a;\theta_a)$ w.r.t. action $a$. 

By randomly sampling $N_{batch}$ historical transitions from the experience replay buffer ${\bm R}$,
$\nabla_{\theta_a} J(\theta_a)$ is approximated by
\begin{equation}\label{eq-gradient-approx}
\nabla_{\theta_a} J(\theta_a) 
\!\approx\! \frac{1}{N_{batch}} \!\sum_{i = 1}^{N_{batch}}\!\left[\!\nabla_{\theta_a} \mu(s_i) \nabla_{a} {{Q}}_{\mu}(s_i,a;\theta_c)|_{a = \mu(s_i)} \!\right].
\end{equation}
The parameter of the actor, i.e., $\theta_a$, is updated by using the gradient descent method~\cite{sutton1999policy}
\begin{equation}\label{eq-policy-ascent}
\begin{aligned}
\theta_a & \leftarrow \theta_a + \eta_a \nabla_{\theta_a} J(\theta_a)\theta_a \\
& \quad+ \frac{\eta_a}{N_{batch}} \!\sum_{i = 1}^{N_{batch}}\!\left[\nabla_{\theta_a} \mu(s_i) \nabla_{a} {{Q}}_{\mu}(s_i,a;\theta_c)|_{a = \mu(s_i)} \!\right],
\end{aligned}
\end{equation} 
where $\eta_a$ is the learning rate of the actor network.

One issue of deterministic policies is that they can cause overfitting and shrink the peaks used to produce Q-value estimates~\cite{Silver2014ddpg}.
Specifically, when updating the critics in the DDPG model, the target Q-value produced by the deterministic policies are susceptible to the inaccuracies caused by the Q-function estimation errors. 
\emph{Target policy smoothing}, a regularization strategy for Q-function value learning~\cite{fujimoto2018addressing}, is used to reduce the inaccuracies. 

Based on the randomly sampled $N_{batch}$ past transitions from the experience replay buffer ${\bm R}$, the target action after target policy smoothing is given by
\begin{equation}
a'_{t} = {\text{clip}}\Big( \mu'\left(s_{t+1}; \theta'_a \right) + {\text{clip}\left( \epsilon', -\sigma_m^2, \sigma_m^2\right)},
a_{\min}, a_{\max}\Big), 
\end{equation}
where the noise $\epsilon'$ is randomly sampled from a zero-mean GN
with variance $\sigma_a^2$, i.e., $\epsilon' \sim {\cal N}(0,\sigma_a^2)$; 
and $\sigma_m^2$ is the maximum exploration noise supported by the environment.
The mean square error (MSE)-based losses coming from the two critics are evaluated as
\begin{equation}\label{eq-loss}
\begin{aligned}
L_k(\theta_k) & = \mathbb{E}_{s_t \sim \rho^{\mu}, a_t \sim \mu(s_t;{\theta_a})}\left[\left( Q_{k}\left(s_t, a_t;\theta_k \right) - y_t \right)^2\right],
\end{aligned}	
\end{equation}
where $k= 1$ or $2$; $y_t = r_t + \gamma \min \{Q'_{1}(s_{t+1},a'_{t};\theta'_1),$ $Q'_{2}(s_{t+1},a'_{t};\theta'_2) \}$
is the target Q-value produced by the two target-critics based on the current transition $(s_t, a_t,r_t, s_{t+1})$;
and $\theta'_1$ and $\theta'_2$ are decayed copies of $\theta_1$ and $\theta_2$, respectively.
The smaller Q-value produced by the two target-critics is taken as the target Q-value.

With $N_{batch}$ randomly sampled transitions, the loss function, $L_k(\theta_k)$, is approximated by
\begin{equation}\label{eq-loss 2}
\begin{aligned}
L_k(\theta_k) \approx \frac{1}{N_{batch}} \sum_{i = 1}^{N_{batch}}\left[\left( Q_{k}\left(s_i, a_i\right) - y_i \right)^2\right], \; k= 1,2,
\end{aligned}
\end{equation}
where $y_i = r_i + \gamma \min\Big(Q'_{1}(s_{i+1},a'_{i};\theta'_1), Q'_{2}(s_{i+1},a'_{i};\theta'_2) \Big)$ is the approximate target Q-value produced by the target network based on the $N_{batch}$ randomly sampled transitions. 
The smaller approximate target Q-value produced by the two target-critics is taken as the approximate target Q-value.

By differentiating $L_k(\theta_k)$ w.r.t. $\theta_k$, we obtain the gradient as
\begin{equation}\label{eq-critic-update}
\begin{aligned}
\nabla_{\theta_k} L_k(\theta_k) 
&\!\approx\! \frac{1}{N_{batch}} \!\sum_{i = 1}^{N_{batch}}\!\Big[\left( Q_{\mu}\left(s_i, \mu(s_i; \theta_a);\theta_k \right) - y_i \right)  \\
&\quad\nabla_{\theta_k} {{Q}}_{\mu}(s_i,\mu(s_i; \theta_a);\theta_k)\Big], \; k= 1,2. 
\end{aligned}	
\end{equation}
The parameters of the two critics, i.e., $\theta_1$ and $\theta_2$, are updated by utilizing the stochastic gradient descent method~\cite{sutton1999policy}. 

According to the \emph{``delayed'' policy updates},
the target-actor and the two target-critics evolving from the actor and critics are updated every two iterations by running the Polyak Averaging~\cite{dankwa2019twin}:
\begin{equation}\label{eq-Polyak}
\begin{aligned}
\theta'_a &\leftarrow \rho_{\tau} \theta_a + (1-\rho_{\tau})\theta'_a,\;\\
\theta'_k &\leftarrow \rho_{\tau} \theta_k + (1-\rho_{\tau})\theta'_k,\; k =1,2,
\end{aligned}	
\end{equation} 
where $\rho_{\tau}$ is the decaying rate of both the actor and critic networks.

\subsection{Optimal Channel Allocation and Rate Adaptation}\label{subsec-optimization}
Given the reflection matrix of the RIS, $\Phit$, from the TD3 network, the effective channel gains of the BS and jammer to the $m$-th user in the $k$-th subchannel, $|h_{m,k}|^2 $ and $|h^J_{m,k}|^2 $, are readily measurable. We can rewrite problem \textbf{P1} as 
\begin{equation}\label{eq-P2}
	\textbf{P2}:\;\max_{\bm{\eta}}\;\;  R_{\rm{tot}}\left({\bm{\eta}} \right),\;\;
	{\text{s.t.}}\; 
	 \eqref{eq-P1 b},\,\eqref{eq-P1 d} - \eqref{eq-P1 g}.
\end{equation} 
By defining $\lambda$ as the Lagrange multiplier w.r.t~\eqref{eq-P1 b}, 
and ${\bm{\nu}} = \{\nu_m,\forall m\}$ as the Lagrange multipliers w.r.t \eqref{eq-P1 g}, 
the Lagrange function of~\eqref{eq-P2} is 
	\begin{equation}\label{eq-lagrange}
	L\left(\bm{\eta}, \lambda, {\bm{\nu}} \right) \!=\! R_{\rm{tot}} \left(\bm{\eta}\right) - \lambda \left( P\left({\bm{\eta}}\right) - P_{\max} \right) -\! \sum_{m=1}^{M}\!\nu_m\left(\!R^{(1)}_m\left(\bm{\eta}\right) - \! \chi R^{(2)}_m\left(\bm{\eta}\right)\!\right) .
	\end{equation}
	Further define
	\begin{equation}\label{eq-varpi}
	\varpi^{(q)}_{m,k,l}\left(\lambda,\nu_m, |h_{m,k}|^2,|h^J_{m,k}|^2 \right) = \left\{
	\begin{aligned}
	&  - \lambda p^{(q)}_{m,k,l}\left(|h_{m,k}|^2,|h^J_{m,k}|^2 \right) \\
	& \;\; + \left(1- \nu_m \right) r_l, {\text{if}}\,q=1;\\
	&  -\lambda p^{(q)}_{m,k,l}\left(|h_{m,k}|^2,|h^J_{m,k}|^2 \right)\\
	& \;\; + \left(1+ \nu_m \chi \right) r_l,{\text{if}}\,q=2.
	\end{aligned}
	\right.
	\end{equation}
	Then, \eqref{eq-lagrange} is rewritten as
	\begin{equation}
	\begin{aligned}
	&L\left(\bm{\eta}, \lambda, {\bm{\nu}} \right)
	= \lambda P_{\max} \!+\! \sum_{k = 1}^K \!\Bigg\lbrace \!\sum_{m = 1}^M \sum_{l=0}^{L}\sum_{q=1}^{Q}\! \eta^{(q)}_{m,k,l}\left(|h_{m,k}|^2, |h^J_{m,k}|^2\right) \\
	&  \qquad\qquad\qquad\qquad\qquad \times\varpi^{(q)}_{m,k,l}\left(\lambda,\nu_m, |h_{m,k}|^2, |h^J_{m,k}|^2 \right)\!\Bigg\rbrace\!.
	\end{aligned}
	\end{equation}
	The Lagrange dual function is
	\begin{equation}\label{eq-lagrange dual}
	D\left(\lambda \right) = \max_{\bm{\eta}} L\left(\bm{\eta}, \lambda, {\bm{\nu}}\right).
	\end{equation}
	The dual problem of \eqref{eq-P2} is given by 
	\begin{equation}\label{eq-dual}
	\min_{\lambda, {\bm{\nu}}} D\left(\lambda, {\bm{\nu}} \right).
	\end{equation}
	Given $\lambda$ and ${\bm{\nu}}$, the primary variable $\bm{\eta} $ is obtained by solving 
	\begin{equation}
	\begin{aligned}
	\max_{\bm{\eta}}\! & \sum_{k = 1}^K \!\Bigg\lbrace \!\sum_{m = 1}^M \sum_{l=0}^{L} \sum_{q=1}^{Q}\! \eta^{(q)}_{m,k,l}\left(|h_{m,k}|^2,|h^J_{m,k}|^2 \right)\\ 
	&\quad\; \times\varpi^{(q)}_{m,k,l}\left(\lambda,\nu_m, |h_{m,k}|^2, |h^J_{m,k}|^2 \right)\!\Bigg\rbrace,\,{\text{s.t.}} \,\eqref{eq-P1 e},~\eqref{eq-P1 f}. 
	\end{aligned}	
	\end{equation}
	The optimal channel allocation and modulation-coding selection take a ``winner-takes-all'' strategy~\cite{He2014Optimal}. As per the $k$-th subchannel, the $m_k^*$-th user and the $l_k^*$-th modulation-coding mode are selected to deliver the $q^{\ast}$-th data stream: 
	\begin{equation}\label{eq-pair}
	\left\lbrace m^{\ast}_k, l^{\ast}_k, q^{\ast}_k \right\rbrace = \arg \max_{m,l,q}\!\; \varpi^{(q)}_{m,k,l}\left(\!\lambda,\nu_m, |h_{m,k}|^2,|h^J_{m,k}|^2\!\right),\; \forall k \in {\cal K}.
	\end{equation}	
	A greedy strategy can be taken to optimize $\bm{\eta}$:
	\begin{equation}\label{eq-opt-eta}
	\left\lbrace
	\begin{aligned}
	&\eta^{(q)\ast}_{m,k,l} \left(\! \lambda,\nu_m ,|h_{m,k}|^2, |h^J_{m,k}|^2\!\right) = 1,\, {\rm{if}}\, \left\lbrace m, l, q \right\rbrace = \left\lbrace m^{\ast}_k, l^{\ast}_k, q^{\ast}_k \right\rbrace;\\
	&\eta^{(q)\ast}_{m,k,l} \left(\! \lambda,\nu_m ,|h_{m,k}|^2, |h^J_{m,k}|^2\!\right) = 0,\, {\rm{otherwise}}.
	\end{aligned}\right.	 
	\end{equation}
	
	With ${\bm{\eta}^{\ast}\left(\lambda, {\bm{\nu}} \right)}$ obtained in \eqref{eq-opt-eta}, the sub-gradient descent method is taken to update $\lambda$ and ${\bm{\nu}}$ by solving the dual problem~\eqref{eq-dual}. 
	$\lambda$ and ${\bm{\nu}}$ are updated by~\cite{boyd2004convex}
	\begin{subequations}\label{eq-sub-gradient}
		\begin{align}
		\lambda\left(\tau +1 \right) & = \left[ \lambda(\tau ) + \varepsilon\left(P\left({\bm{\eta}}^{\ast} \left(\lambda\left(\tau \right),{\bm{\nu}}(\tau ) \right) \right) - P_{\max}\right) \right]^+,\\
		\nu_m\left(\tau +1 \right) 
		& = \left[ \nu_m (\tau ) + \varepsilon\left(R^{(1)}_m \left({\bm{\eta}}^{\ast} \left(\lambda\left(\tau \right),{\bm{\nu}}(\tau ) \right)\right) \right.\right.\notag\\
		&\left.\left.\qquad\quad\;\,- \chi R^{(2)}_m \left({\bm{\eta}}^{\ast} \left(\lambda\left(\tau \right),{\bm{\nu}}(\tau ) \right)\right)\right) \right]^+,\; \forall m,
		\end{align}	
	\end{subequations}
	where $\varepsilon$ is the step size, $\tau$ is the index to the iterations, and $\left[x \right]^+ = \max\left(0,x \right)$. At initialization, $\lambda$ and $\bm{\nu}$ are non-negative, i.e., $\lambda(0) \geq 0$ and $\nu_m(0) \geq 0, \forall m$, to ensure \eqref{eq-sub-gradient} converges.
	
	It is prudent to analyze the optimality of the solution obtained iteratively by~\eqref{eq-opt-eta} and \eqref{eq-sub-gradient}, since problem \eqref{eq-P2} is a non-convex mixed-integer program. 
	We assert that when the gains of the channels, $|h_{m,k}|^2$ and $|h^J_{m,k}|^2, \forall m\in {\cal M},\, k\in {\cal K}$, have a continuous cumulative distribution function (CDF).
	$\eta^{(q)\ast}_{m,k,l} \left( \lambda^{\ast},\nu^{\ast}_m,|h_{m,k}|^2, |h^J_{m,k}|^2\right), \forall m,k$, is the almost surely optimal solution to problem \textbf{P2} (i.e., with probability 1), where $\lambda^{\ast}$ is obtained in \eqref{eq-sub-gradient} with any initial $\lambda(0) > 0$ and $\nu_m(0) > 0$.
	The proof can be referred to~\cite{He2014Optimal}. For the completeness of this paper, a sketch of the proof is provided below. 

	The proof starts by confirming the almost sure uniqueness of the ``winner-takes-all'' strategy ${\bm{\eta}}^{\ast}(\lambda, {\bm{\nu}})$ in all three possible cases. $(a)$ If $\max_{m,l,q} \varpi^{(q)}_{m,k,l}\left(\lambda, \nu_m, |h_{m,k}|^2, |h^J_{m,k}|^2\right) =0$, all users undergo a deep fade in the $k$-th subchannel.
	Even if user $m$ is selected for the subchannel, $l^{\ast}_k(\lambda, \nu_m, |h_{m,k}|^2, |h^J_{m,k}|^2) = 0$, the optimal decision of the BS is to not transmit in the subchannel; see~\eqref{eq-opt-eta}. 
	$(b)$ If $\max_{m,l,q} \varpi^{(q)}_{m,k,l}\left(\lambda,\nu_m, |h_{m,k}|^2, |h^J_{m,k}|^2\right) > 0$ and a single ``winner'' wins the $k$-th subchannel, the optimal strategy in \eqref{eq-opt-eta} is unique. 
	$(c)$ If $\max_{m,l,q} \varpi^{(q)}_{m,k,l}\left(\lambda, \nu_m, |h_{m,k}|^2, |h^J_{m,k}|^2 \right) > 0$ and multiple $\{m,l,q\}$ triplets can win the $k$-th subchannel with one triplet selected at random, the strategy is non-unique. 
	This is a Lebesgue measure zero event~\cite{lenz2002singular} under the continuous CDF of the random channel gain. 
	The non-unique ``winner'' has the ``measure zero'' effect, i.e., the probability of the non-unique ``winner'' is almost zero.
    Given its almost sure uniqueness, the ``winner-takes-all" strategy maximizes the Lagrangian function \eqref{eq-lagrange dual}, even if~\textbf{P2} is relaxed to a linear program (LP), i.e., ${\bm{\eta}}^{\ast}(\lambda, {\bm{\nu}})$ can take a continuous value within $[0,1]$. Since the LP has a zero-duality gap, 
    ${\bm{\eta}}^{\ast}(\lambda, {\bm{\nu}})$ is almost surely optimal for~\textbf{P2}.	 

\begin{algorithm}[t]\small
	\DontPrintSemicolon
	\caption{Proposed PSD-TD3 to solve problem~\textbf{P1}.}
	\label{algo_TD3}
	{\bf Initialization:}
	Randomly initialize the actor $\mu$ and the two critics ${Q}_{1}$ and  ${Q}_{2}$ with parameters $\theta_a$, $\theta_1$, and $\theta_2$, the target-actor $\mu'$ and two target-critics ${Q}'_{1}$ and ${Q}'_{2}$ with parameters $\theta'_a \leftarrow \theta_a$, $\theta'_1 \leftarrow \theta_1$, and $\theta'_2 \leftarrow \theta_2$, the experience replay buffer ${\bm{R}}$, and the channel allocation and modulation-coding selection~${\bm \eta}_0$.\\
	Measure the received data rates of all users and ${\bm \eta}_0$ as the initial state $s_0$.\\
		\For{$t= 1,\cdots,T_s$}{
			Pick action $a_t = {\text{clip}}\Big(\mu\left(s_t; \theta_a \right) + \epsilon, a_{\min}, a_{\max} \Big)$, and update $\Phit$.\\
			Obtain the dual problem of \textbf{P2} based on the updated $\Phit$: $\min_{\lambda, {\bm{\nu}}}\max_{\bm{\eta}} L\left(\bm{\eta}, \lambda, {\bm{\nu}} \right)$.\\
			Initialize $I = 0$, the maximum iteration number $I_{\max}$, $\lambda (0) \geq 0$, $\nu_m (0) \geq 0, \forall m$, and $\bm{\eta}_0$.\\
			\While{$L\left(\bm{\eta}, \lambda, {\bm{\nu}} \right)$ is yet to converge, and $I < I_{\max}$}{
			Obtain $\bm{\eta}^\ast$ by maximizing $L\left(\bm{\eta}, \lambda, {\bm{\nu}} \right)$ given $\lambda$ using a greedy strategy.\\
			Initialize $J = 0$ and the maximum iteration number $J_{\max}$: \\
			\While{$P\left({\bm{\eta}^\ast}\right)$ is yet to converge, and $J < J_{\max}$}{
			Update $\lambda$ and $\nu_m, \forall m$ according to \eqref{eq-sub-gradient}.\\
					$J \leftarrow J + 1$.}		
					$I \leftarrow I + 1$.
				}    
				Output the optimal channel allocation and modulation-coding selection ${\bm \eta}_t = \bm{\eta}^{\ast}$.\\ 
			Receive the reward $r_t$, perceive a new state $s_{t+1}$, and reserve transition $(s_t, a_t, r_t, s_{t+1})$ in ${\bm{R}}$.\\
			Randomly sample $N_{batch}$ historical transitions $(s_i, a_i, r_i, s_{i+1})$ from ${\bm{R}}$.\\
			Update the target action after target policy smoothing based on the sampled transitions:
			$a'_{t} = {\text{clip}}\Big( \mu'\left(s_{t+1}; \theta'_a \right) + {\text{clip}\left( \epsilon', -\sigma_m^2, \sigma_m^2\right)},
			a_{\min}, a_{\max}\Big)$.\\
			Update the target Q-value: $y_i = r_i + \gamma \min\Big(Q'_{1}(s_{i+1},a'_{i};\theta'_1), Q'_{2}(s_{i+1},a'_{i};\theta'_2) \Big)$.\\
			Calculate the loss function based on~\eqref{eq-loss 2}, and
			update the two critics by~\eqref{eq-critic-update}.\\
			\If{$\mod(t,2) = 0$}{Update the actor based on~\eqref{eq-policy-ascent}, and the target-actor and the two target-critics by~\eqref{eq-Polyak}.\\
			}   
		}
\end{algorithm}
\subsection{Algorithm Description}
Algorithm~\ref{algo_TD3} summarizes the proposed algorithm, referred to as PSD-TD3.
The agent collects the received data rates of the users
at the start of every learning step (i.e., the $t$-th step), and takes them as the state of the algorithm (i.e., state $s_t$) to train the actor. A continuous action $a_t$ is produced by the actor to update the reflection matrix of the RIS using TD3; see Section~\ref{subsec-ddpg}.
Given the reflection matrix, the algorithm optimizes the channel and modulation-coding selection, i.e.,~$\bm{\eta}_t$, using PSD; see Section~\ref{subsec-optimization}. 
Based on the selection, 
the agent evaluates the reward~$r_t$, transits to the state $s_{t+1}$, and reserves transition $(s_t, a_t, r_t, s_{t+1})$ in the experience replay buffer ${\bm R}$.
The parameters of the six DNNs are updated with randomly sampled past transitions in the experience replay buffer ${\bm R}$ until the cumulative reward converges. 

It is noted that our proposed framework can be readily extended to a multi-antenna setting where 
both the BS and users can have multiple antennas. 
In this case, space-time block coding (STBC) and maximal ratio combining (MRC) can be carried out at the BS and users, respectively. Given an RIS configuration, each user can individually measure its effective channel matrix from the BS, denoted by {${\Ht}_{m,k} = \left(\Ht^{br}_k \right)^H \Phit^H \Ht^{ru}_{m,k} + \Ht^d_{m,k}\in \mathbb{C}^{N_t \times N_r}, \; \forall m \in {\cal M},\; \forall k \in {\cal K}$}, and evaluate and report its effective channel gain of each subchannel, i.e., 
$\gamma_{m,k}^{\rm{mrc}} = \frac{p_{m,k} \left(\| {\Ht}_{m,k}\|^2/ N_t N_r R_c\right)}{p^J_{k}(\|\hf^J_{m,k}\|^2/N_r R_c) + \sigma^2}$~\cite{Yang2021Intelligent,Yang2021Deep,2004MaarefAdaptive}, where $R_c$ is the information code rate of the STBC, and $N_t$ and $N_r$ are the numbers of antennas at the BS and users, respectively. Accordingly, the BS can optimize the selections of the user, data stream (with a quality requirement), and modulation-coding scheme, and the allocation of its transmit power for each subchannel in the same way as it does under the single-antenna setting.

\subsection{Complexity and Convergence Analyses}
The computational complexity of the proposed PSD-TD3 algorithm accounts for both the PSD and the TD3 model.
	As of the PSD, the BS incurs a linear complexity of $O(KMLQ)$ to evaluate the net rewards 
	$\varpi_{m,k,l}^{(q)}\left(\lambda, \nu_m, |h_{m,k}|^2, |h^{J}_{m,k}|^2 \right)$ for all $M$ users with $Q$ data streams per user, $K$ channels, and $L$ modulation-coding modes. Moreover, the greedy strategy, i.e.,~\eqref{eq-opt-eta}, used to decide the 3-tuple  $\left\lbrace m^{\ast}_k, l^{\ast}_k, q^{\ast}_k \right\rbrace$ 
 per subchannel $k$ incurs a complexity of ${\cal O}(K\log(MLQ))$. As a result, all required operations of a  time step incurs a linear
	computational complexity of ${\cal O}(KMLQ)$~\cite{He2014Optimal}. 
	
    As of the TD3 model, we separately evaluate the complexities of  the actor and critic networks.
Suppose that the actor network has $L_a$ layers with $J_m$ neurons in the $m$-th layer ($m\leq L_a$). The complexity of the $m$-th layer is ${\cal O}(J_{m-1}J_m + J_mJ_{m+1})$~\cite{Gao2020DeepComfort}. The complexity of the actor network is ${\cal O}\left(\sum_{m=2}^{L_a-1}(J_{m-1}J_m + J_mJ_{m+1})\right)$.
Suppose that the critic network has $L_c$ layers with  $G_n$ neurons in the $n$-th layer ($n\leq L_c$). The  complexity of the $n$-th layer is ${\cal O}(G_{n-1}G_n + G_nG_{n+1})$~\cite{Gao2020DeepComfort}. The complexity of the critic network is ${\cal O}\left(\sum_{n=2}^{L_c-1}(G_{n-1}G_n + G_nG_{n+1})\right) $.
	As a result, the overall computational complexity of the TD3 model is ${\cal O}\left( \sum_{m=2}^{L_a-1}(J_{m-1}J_m + J_mJ_{m+1}) + \sum_{n=2}^{L_c-1}(G_{n-1}G_n + G_nG_{n+1})\right) $~\cite{Gao2020DeepComfort}.

We further analyze the convergence of the proposed PSD-TD3 algorithm.
Specifically, the algorithm satisfies the following conditions:
(i) the network parameters $\theta$ and $\theta'$ (of which the subscripts are suppressed for brevity) are upper bounded since they are sequentially compact following the Arzela-Ascoli theorem \cite{billingsley2013convergence};
(ii) the state and action spaces are compact as the sampled states and actions are bounded by the maximum transmit power of the BS and the phase shifts of the RIS;
(iii) the reward function, i.e., \eqref{eq-reward}, is continuous;
and (iv) the training networks are feedforward FCNNs with twice continuously differentiable activation functions, such as Rectified Linear Units (ReLUs) and sigmoid.
According to \cite[Lemma~2]{redder2022asymptotic}, the proposed algorithm can asymptotically converge 
if we adopt a sequence of square summable learning rates, i.e., $\sum_t \eta_a (t) = \infty$
and $\sum_t \eta_a (t)^2 < \infty$. Here, $t$ is the time step, and $\eta_a (t)$ is a time-varying learning rate of the actor network.

\section{Simulation Results}\label{sec-sim}	
In the considered system, the BS is placed at $(D_0,0,H_b)$, the jammer is placed at $(x_J,y_J,0)$, and the first element of the RIS has the coordinates $(0,\delta,\delta+H_r)$, as depicted in Fig.~\ref{fig-sysmodel}. 
We set $D_0 = 2$ m, $H_b = 10$ m, $H_r =10$ m, $x_J =50$ m, and $y_J = 150$ m.
The RIS is a URA with element spacing of $\delta$. We assume $d_0 = \delta = \frac{\lambda}{2}$. 
We use $(\iota,\kappa)$ to index the RIS elements. $\iota \in \{1,\cdots,{N_y}\}\;{\text{and}}\;\kappa \in \{1,\cdots,{N_z}\}$. 
The coordinates of the $(\iota,\kappa)$-th reflecting element are $(0,\iota\times\delta,\kappa\times\delta+H_r)$.
The users are uniformly scattered within a square area centered at $(100, 100, 0)$~m with the side length of 100~m. 
The sides of the area are parallel to the $x$- and $y$-axes. 
The location of the $m$-th user is $(x_{m},y_{m},0)$, $\forall {m} \in {\cal M}$.
By default, $M=4$. 

We consider Rayleigh fading for the BS-user (BS-UE) and the jammer-UE links, and Rician fading for the BS-RIS, jammer-UE and  RIS-UE links. The channel gains of the BS-UE (or jammer-UE), BS-RIS (or jammer-RIS), and RIS-UE links are given by
\begin{align}\label{eq-bs-ue}
\!h^{d}_{m,k}& =\! \sqrt{\epsilon_o \left(d^{d}_{m} \right)^{-\alpha_{d}}} \tilde{h}^d, \forall m,k,&\\
\!h^{br}_{\iota,\kappa}& = \!\sqrt{\epsilon_o \left(d^{br}_{\iota,\kappa} \right)^{-\alpha_{br}}}\left(\! \sqrt{\frac{K_1}{1\!+\!K_1}} h^{br}_{los} \!+\! \sqrt{\frac{1}{1\!+\!K_1}}h^{br}_{nlos} \!\right), \forall \iota,\kappa, & \label{eq-bs-irs}\\
\!h^{ru}_{\iota,\kappa,m} &=\! \sqrt{\epsilon_o \left(d^{ru}_{\iota,\kappa,m} \right)^{-\alpha_{ru}}}\left(\! \sqrt{\frac{K_2}{1\!+\!K_2}} h^{ru}_{los} \!+\! \sqrt{\frac{1}{1\!+\!K_2}}{h}^{ru}_{nlos} \!\right), \forall \iota,\kappa,m, & \label{eq-irs-ue}\\
\!h^{Jd}_{m,k}& =\! \sqrt{\epsilon_o \left(d^{Jd}_{m} \right)^{-\alpha_{Jd}}} \tilde{h}^{Jd}, \forall m,k,& \label{eq-jammer-ue}\\
\!h^{Jr}_{\iota,\kappa}& = \!\sqrt{\epsilon_o \left(d^{Jr}_{\iota,\kappa} \right)^{-\alpha_{Jr}}}\left(\! \sqrt{\frac{K_3}{1\!+\!K_3}} h^{Jr}_{los} \!+\! \sqrt{\frac{1}{1\!+\!K_3}}h^{Jr}_{nlos} \!\right), \forall \iota,\kappa, & \label{eq-jammer-irs}
\end{align}
where $\epsilon_o$ is the path loss at the reference distance $d_0 =1$ m with $\alpha_{d}$, $\alpha_{br}$, $\alpha_{ru}$, $\alpha_{Jd}$, and $\alpha_{Jr}$ being the path loss exponents of the BS-RIS, BS-UE, RIS-UE, jammer-UE, and jammer-RIS links, respectively; $d^{br}_{\iota,\kappa} = \sqrt{\left(H_r+\kappa\delta -H_b \right)^2+\iota^2\delta^2+D_0^2}$ is the distance from the BS to the $(\iota,\kappa)$-th reflecting element of the RIS, and $d^{d}_{m} = \sqrt{\left(D_0 - x_{m}\right)^2 + y_{m}^2 + H_b^2}$ is the distance from the BS to the $m$-th user, and $d^{ru}_{\iota,\kappa,m} = \sqrt{x_m^2 + \left(H_r + \kappa\delta \right)^2 +(y_m -\iota\delta)^2}$ is the distance from the $(\iota,\kappa)$-th reflecting element of the RIS to the $m$-th user,
$d^{Jr}_{\iota,\kappa} = \sqrt{\left(H_r+\kappa\delta \right)^2+\left(\iota\delta - y_J\right)^2+x_J^2}$ is the distance from the jammer to the $(\iota,\kappa)$-th reflecting element of the RIS, $d^{Jd}_{m} = \sqrt{\left(x_J - x_{m}\right)^2 +\left(y_J - y_{m}\right)^2}$ is the distance from the jammer to the $m$-th user. 

In \eqref{eq-bs-irs},~\eqref{eq-irs-ue}, and \eqref{eq-jammer-irs}, $K_1$, $K_2$ and $K_3$ are the Rician factors of the BS-RIS, RIS-UE, and jammer-RIS links.
$h^{br}_{los} = e^{-j \frac{2\pi \delta}{\lambda}\phi^{br}_{\iota, \kappa}}$, $h^{ru}_{los} = e^{-j \frac{2\pi \delta}{\lambda}\phi^{ru}_{\iota, \kappa, m}}$, and $h^{Jr}_{los} = e^{-j \frac{2\pi \delta}{\lambda}\phi^{Jr}_{\iota, \kappa}}$ are the deterministic Line-of-Sight (LoS) components of the BS-RIS, RIS-UE, and jammer-RIS links, respectively, 
where $\phi^{br}_{\iota, \kappa} = \arccos\left(\frac{\iota \delta}{d^{br}_{\iota, \kappa}} \right) $ is the angle-of-arrival (AoA) of the signal from the BS to the $(\iota, \kappa)$-th reflecting element of the RIS, 
$\phi_{ru} = \arccos\left(\frac{y_m - \iota \delta}{d^{ru}_{\iota,\kappa,m}} \right) $ is the angle-of-departure (AoD) of the signal from the $(\iota, \kappa)$-th reflecting element of the RIS to the $m$-th user,
and $\phi^{Jr}_{\iota, \kappa} = \arccos\left(\frac{y_J - \iota \delta}{d^{Jr}_{\iota, \kappa}} \right) $ is the AoA of the signal from the jammer to the $(\iota, \kappa)$-th reflecting element of the RIS. 
$\tilde{h}^d$, $h^{br}_{nlos}$, $h^{ru}_{nlos}$, $\tilde{h}^{Jd}$, and $h^{Jr}_{nlos}$ are random scattering components modeled by zero-mean and unit-variance CSCG variables.
The other parameters of the considered system are provided in Table \ref{tab.pm}.
\begin{table}[t]
	\caption{The parameters of the considered system}
	\begin{center}
		\begin{tabular}{ll}
			\toprule[1.5pt]
			Parameters  & Values \\ \hline
			Maximum transmit power of the BS, $P_{\max}$    & 5 -- 35 dBm \\ 
			Transmit power of the jammer, $P_J$ & 10 dBm \\
			Number of subchannels, $K$            & 16, 32 \\ 
			Number of users, $M$              & 4 \\ 
			Number of modulation levels, $L$   & 4 \\
			Set of modulation-coding rate & \{0,2,4,6\} bits/symbol\\
			Path loss at $d_0 =1$ m, $\epsilon_o$      &-30 dB \\ 
			Path loss exponents, $\alpha_{br}$, $\alpha_{d}$, $\alpha_{ru}$   & 2.5, 3.0, 2.2 \\ 
			Rician factors, $K_1$, $K_2$, $K_3$          & 1, 3, 1     \\ 
			Noise power density, $\sigma^2$                 & -169 dBm/Hz \\ 
			Bandwidth, $B_w$                         & 100~MHz \\
			BER requirements, $\{\varrho_0^{(1)},\varrho_0^{(2)}\}$  & $\{10^{-6}, 10^{-2}\}$ \\ 
			Coefficients of modulation and coding, $\beta_1$, $\beta_2$     &0.2, -1.6~\cite{Goldsmith1998} \\
			\toprule[1.5pt]
		\end{tabular}
	\end{center}
	\label{tab.pm}
\end{table}

The TD3-based network is implemented by a two-layer feedforward neural network with 128 and 64 hidden nodes in the two layers.
Rectified Linear Units (ReLUs) are used as the activation functions between the layers of the actor and critic networks.
The output layers of the actor  use the sigmoid($\cdot$) to bound the output actions within $[0, 2\pi)$ for the RIS configuration.
The state and action are taken as the input to the first layer of the critic networks.
The learning rates of both the actor and critic networks are $10^{-3}$.
The exploration noise used to train the TD3 actor, and the policy noise used to smooth the target-actor are both generated from the zero-mean GN with variance $0.2$.
The maximum value of the exploration noise is $0.5$. The update frequency of the actor networks is $2$.
The TD3-based network is trained on a server with an Nvidia Tesla P100 SXM2 16GB GPU. 
The network hyperparameters are summarized in Table~\ref{table_hyper_td3}.
\begin{table}[t]
	\renewcommand{\arraystretch}{1.0}
	\caption{The hyperparameters of the TD3-based algorithm}
	\begin{center}
		\begin{tabular}{ll}
			\toprule[1.5pt]
			Parameters  & Values \\
			\hline
			Discounting factor for future reward, $\gamma$ & 0.99 \\
			Learning rate for actor and critic networks, $\eta_a$, $\eta_c$ & $1 \times 10^{-3}$ \\
			Decaying rate for actor and critic networks, $\rho_{\tau}$ & $5 \times 10^{-3}$\\
			Size of experience replay buffer & $1 \times 10^5$\\
			Number of episodes, $T_{ep}$ & 400 \\
			Total number of steps in each episode, $T_s$ & 200 \\
			Mini-batch size, $N_{batch}$ & 16\\	
			Policy delay update frequency & 2	\\	 
			Maximum value of the Gaussian noise, $\sigma_m^2$ & 0.5 \\
			Variance of the exploration noise, $\sigma_e^2$ & 0.2 \\	
			Variance of the policy noise, $\sigma_a^2$ & 0.2 \\
			\toprule[1.5pt]
		\end{tabular}
	\end{center} \label{table_hyper_td3}	
\end{table}

As discussed earlier, 
no existing algorithm is directly comparable to the proposed PSD-TD3 algorithm. 
We come up with a DDPG-based alternative to the PSD-TD3 algorithm, referred to as PSD-DDPG, where the DDPG is employed to configure the RIS. We also develop a DQN-TD3 algorithm, where the selections of the user, subchannel, and modulation-coding mode are done using a DQN, and the TD3 is used to configure the RIS. 
Moreover, we consider the case where the RIS is randomly configured, while the selections of the user, data stream, subchannel, and modulation-coding mode are optimized, as described in Section~\ref{subsec-optimization}. 
These three benchmarks are used to evaluate the proposed PSD-TDS algorithm.

\begin{figure}[!t]
	\centering
	\includegraphics[width=1\columnwidth]{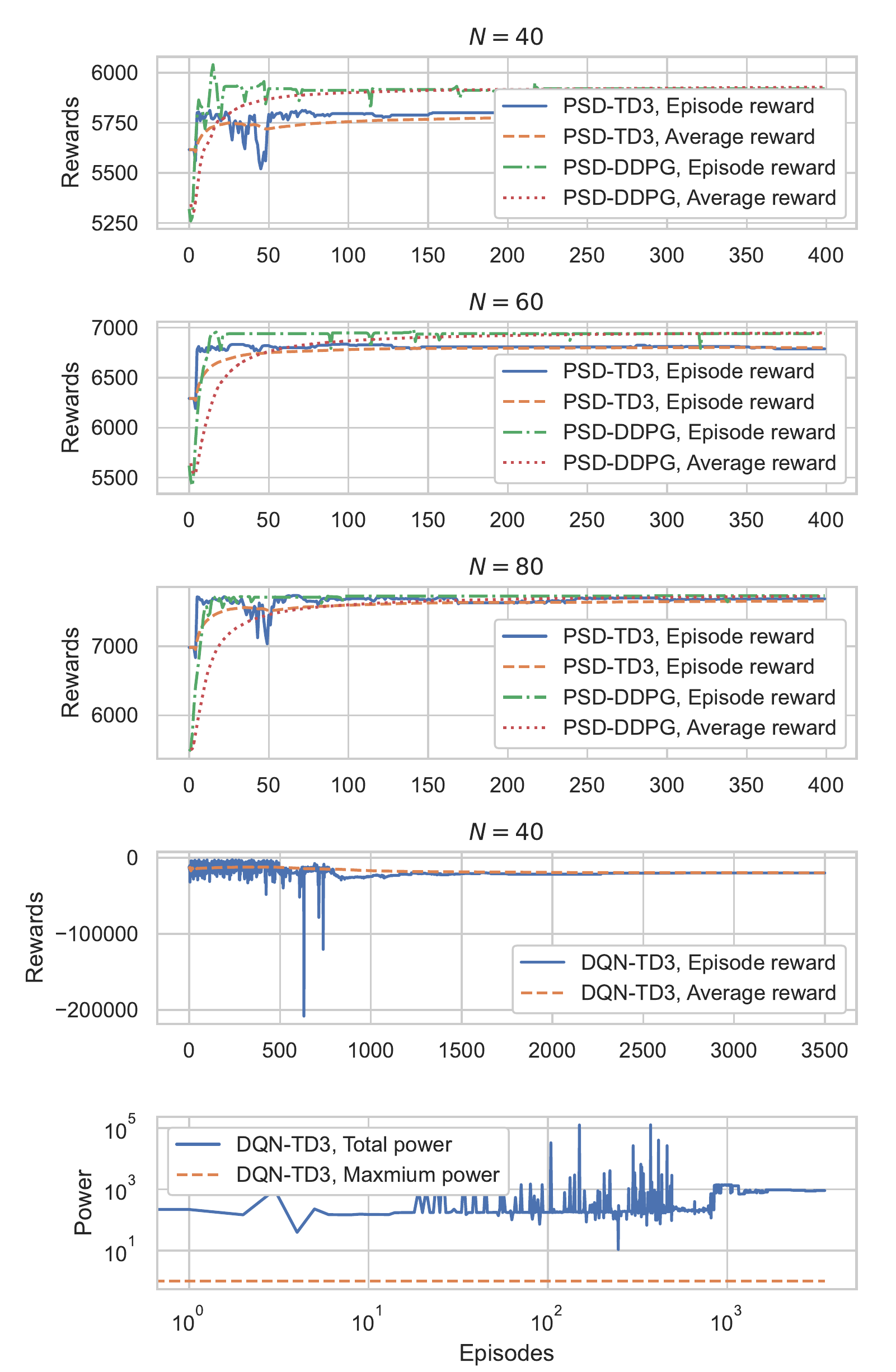}
	\caption{The per-episode and average rewards of the proposed PSD-TD3 algorithm and its DDPG-based alternative under $N=40$, 60, and 80 (the top three subfigures), and the rewards and the BS transmit power of the DQN-TD3 algorithm under $N=40$ (the bottom two subfigures).}
	\label{fig-reward}
\end{figure}

We train the proposed algorithm only for one value of the maximum BS transmit power $P_{\max}$, i.e., $P_{\max} =30$ dBm, and test the resulting model under other $P_{\max}$ values to show the generalizability of the algorithm. Likewise, we train the algorithm only for one value of the transmit power of the jammer $P_J$, and test it under other $P_J$ values.
In the top three subfigures of Fig.~\ref{fig-reward}, we plot both the per-episode reward and the average reward of the proposed PSD-TD3 under different~$N$. 
We also plot the per-episode reward and the
average reward of the alternative PSD-DDPG algorithm.
The average reward over the $i$-th training episode is $\bar r = \frac{1}{T_{s}}\sum_{j=t}^{T_{s}} r_t^i$, where $r_t^i$ is the step reward for the episode; see~\eqref{eq-reward}. 
The top three subfigures of Fig.~\ref{fig-reward} show that the rewards of the two algorithms generally improve with the learning steps, and grow with $N$. 
Moreover, the DDPG-based alternative approach also demonstrates its viability, despite DDPG is known to be susceptible to overfitting (compared to TD3). The conclusion drawn is that the small action space of the new framework, resulting from the decoupling of the discrete and continuous actions, allows even the DDPG model to sufficiently exploit the action space and converge fast.
In the bottom two subfigures of Fig.~\ref{fig-reward}, we see that the rewards of the DQN-TD3 algorithm do not converge to a feasible solution even after 3,500 training episodes since the total transmit power at the BS cannot satisfy the maximum power constraints. In contrast, PSD-DDPG and PSD-TD3 converge within a few episodes. The convergent solutions of PSD-DDPG and PSD-TD3 are inherently feasible, since the transmit power is pre-evaluated before an selection of the user, data stream, subchannel, and modulation-coding mode.

\begin{figure}[t]
	\centering
	\includegraphics[width=1\columnwidth]{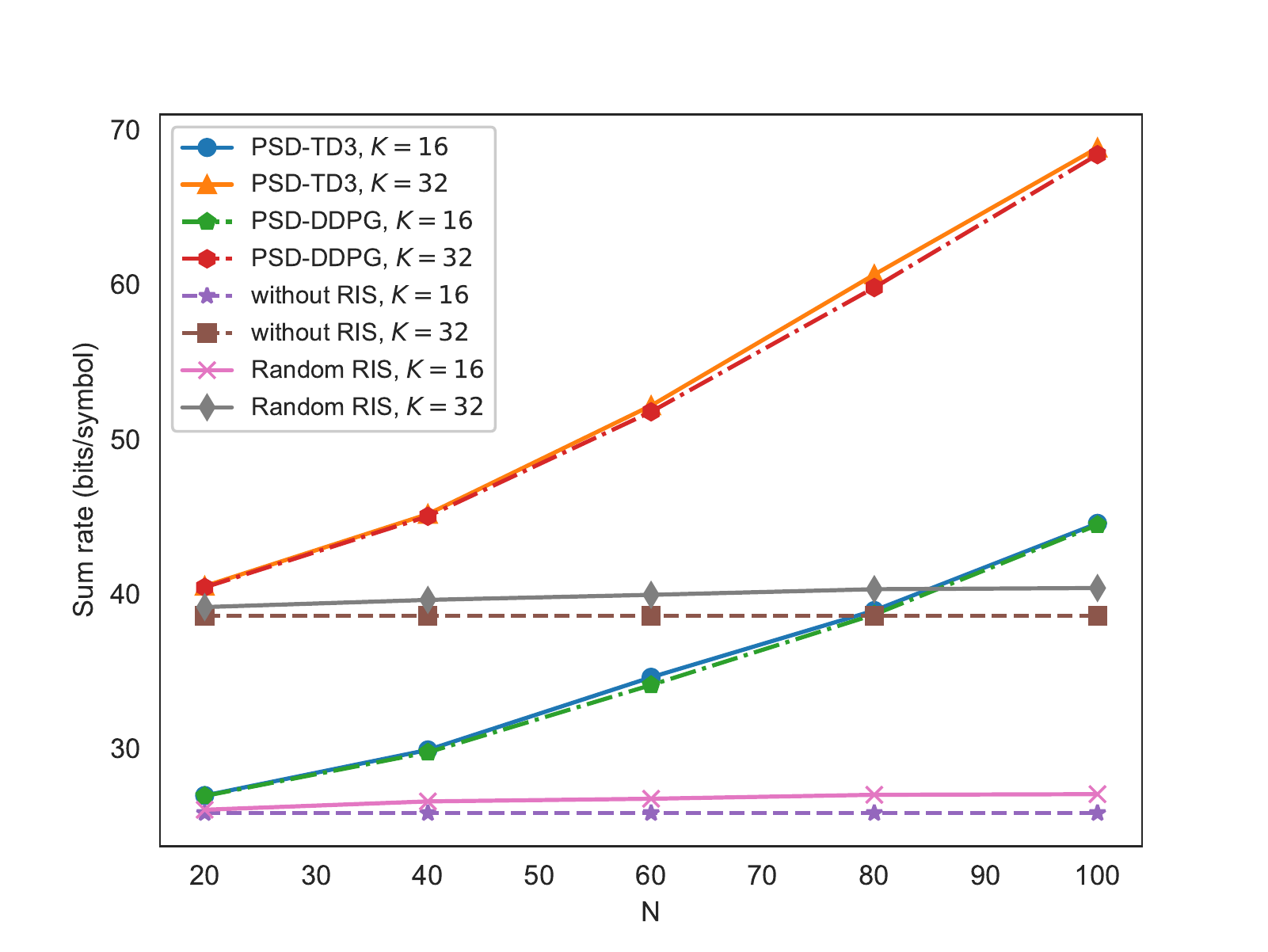}
	\caption{Sum rate vs. the number of RIS's reflecting elements.}
	\label{fig-rate-vs-N}
\end{figure}
\begin{figure}[t]
	\centering
	\includegraphics[width=1\columnwidth]{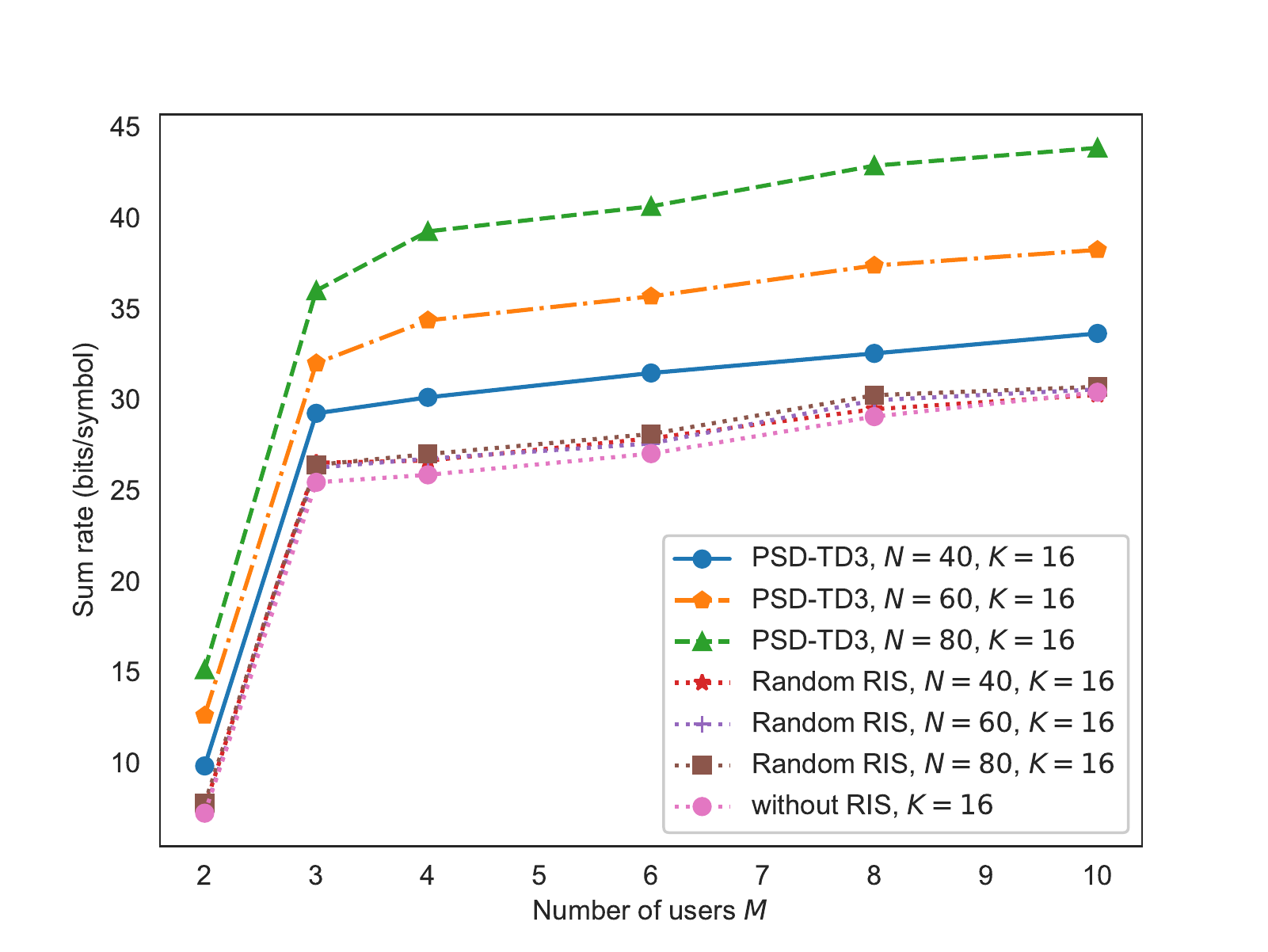}
	\caption{Sum rate vs. $M$, where each value is the average of 200 independent tests.}
	\label{fig_rate_vs_user}
\end{figure}

Next, we examine the proposed PSD-TD3 algorithm and its alternatives under different parameters of the considered system. 
Each testing episode has 200 steps. 
During a testing process, no exploration noise is added.
Fig.~\ref{fig-rate-vs-N} plots the sum rate of the $M$ users against the number of reflecting elements at the RIS, $N$, under $K = 16$ and 32 subchannels. 
We also plot the case with the RIS randomly configured and the case without the RIS for comparison.
We see that both PSD-TD3 and PSD-DDPG are effective and can benefit from the increase of $N$.  
The usefulness of the RIS and the importance of meticulous RIS configuration are demonstrated by comparing the proposed PSD-TD3 to the cases without the RIS and with the RIS randomly configured. 
Particularly, the case with the RIS randomly configured can only marginally outperform the case without the RIS, as will also be shown in Fig.~\ref{fig_rate_vs_user}. 

\begin{figure}[t]
	\centering
	\includegraphics[width=1\columnwidth]{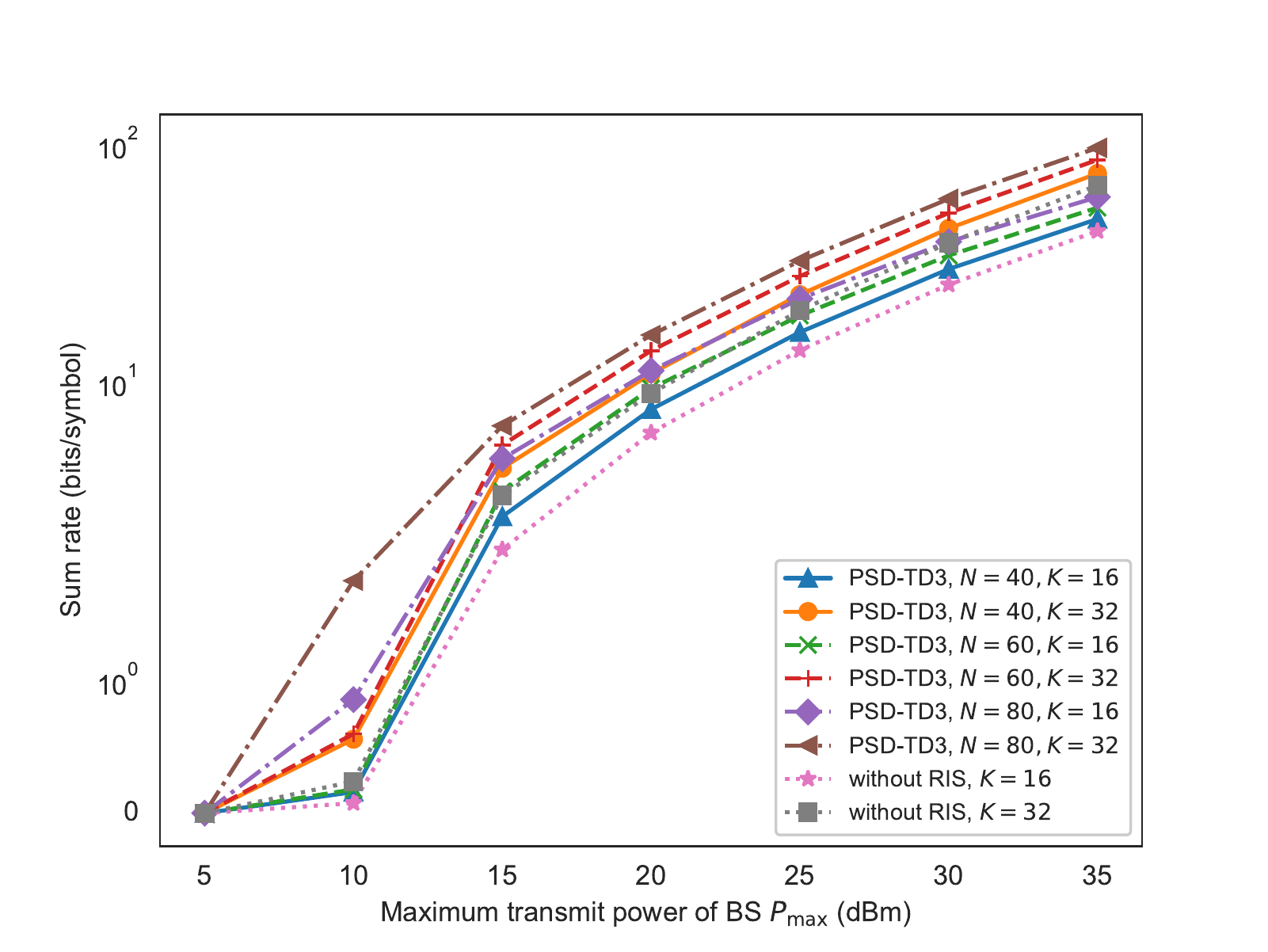}
	\caption{Sum rate vs. $P_{\max}$, where the jamming power is 10~dBm.}
	\label{fig-rate-vs-pmax}
\end{figure}
\begin{figure}[t]
	\centering
	\includegraphics[width=1\columnwidth]{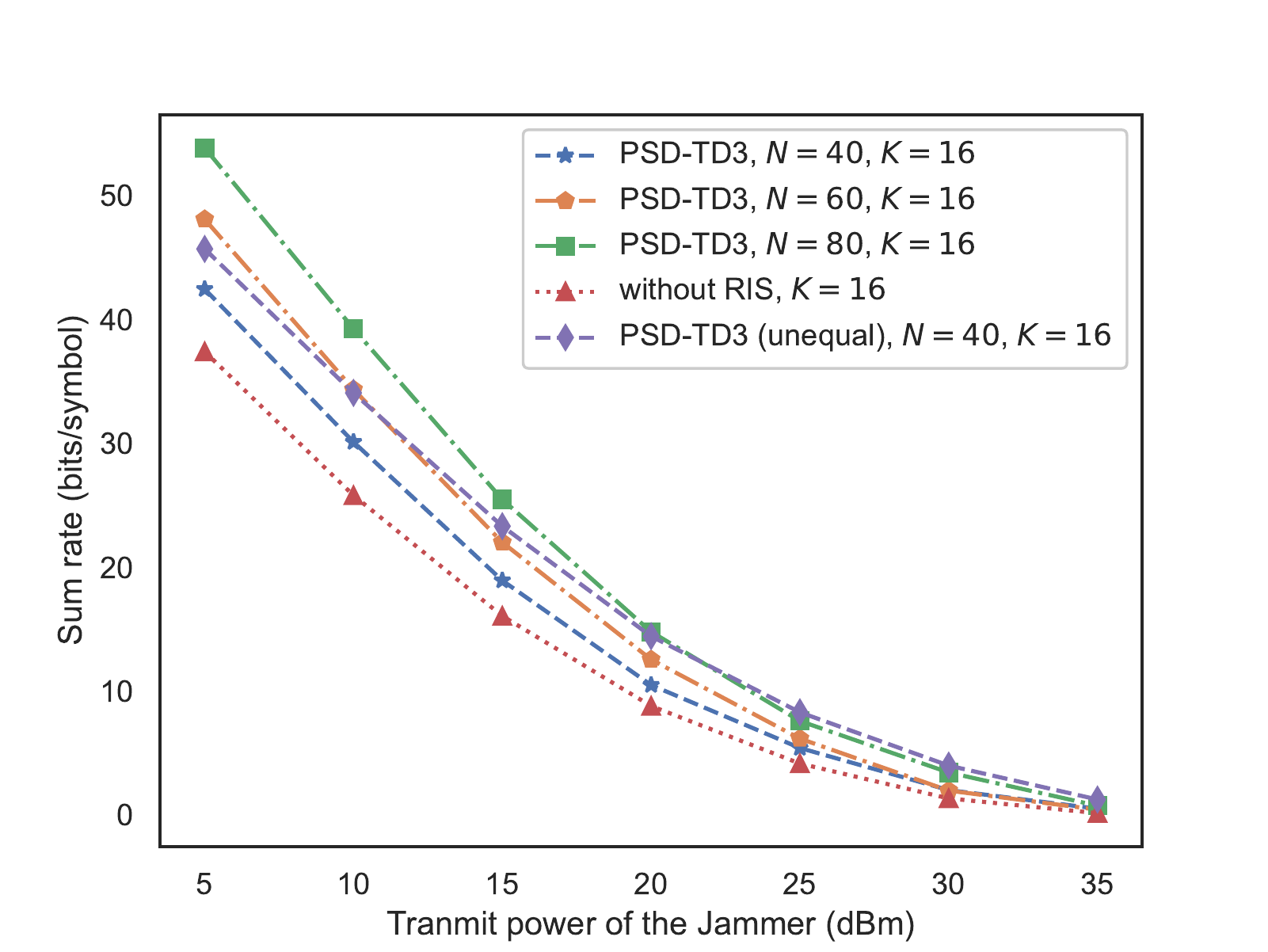}
	\caption{Sum rate vs. the transmit power of Jammer, $P_J$, where the proposed PSD-TD3 is plotted under different sizes of the RIS and compared with the case without the RIS.}
	\label{fig_rate_vs_PJ}
\end{figure}
Fig.~\ref{fig-rate-vs-pmax} plots the sum rate with the increasing maximum transmit power of the BS, $P_{\max}$, under different $N$ and $K$. 
We observe that the proposed PSD-TD3 attains the higher sum rate than the case without the RIS. 
The sum rate grows with $P_{\max}$ under all the considered algorithms and parameter settings. 
The usefulness of the RIS is also validated, since the sum rate grows with $N$.
We also plot the sum rate with the growing transmit power of the jammer, $P_J$, under $K = 16$ in Fig.~\ref{fig_rate_vs_PJ}. We see that the sum rate declines as $P_J$ grows. When $P_J \geq 35$~dBm, the sum rate approaches zero under the proposed PSD-TD3, while it approaches zero when $P_J \geq 25$~dBm in the case without the RIS. In other words, the RIS strengthens the anti-jamming capability significantly by augmenting the radio propagation environment.

\begin{figure}[!t]
	\centering
	\includegraphics[width=1\columnwidth]{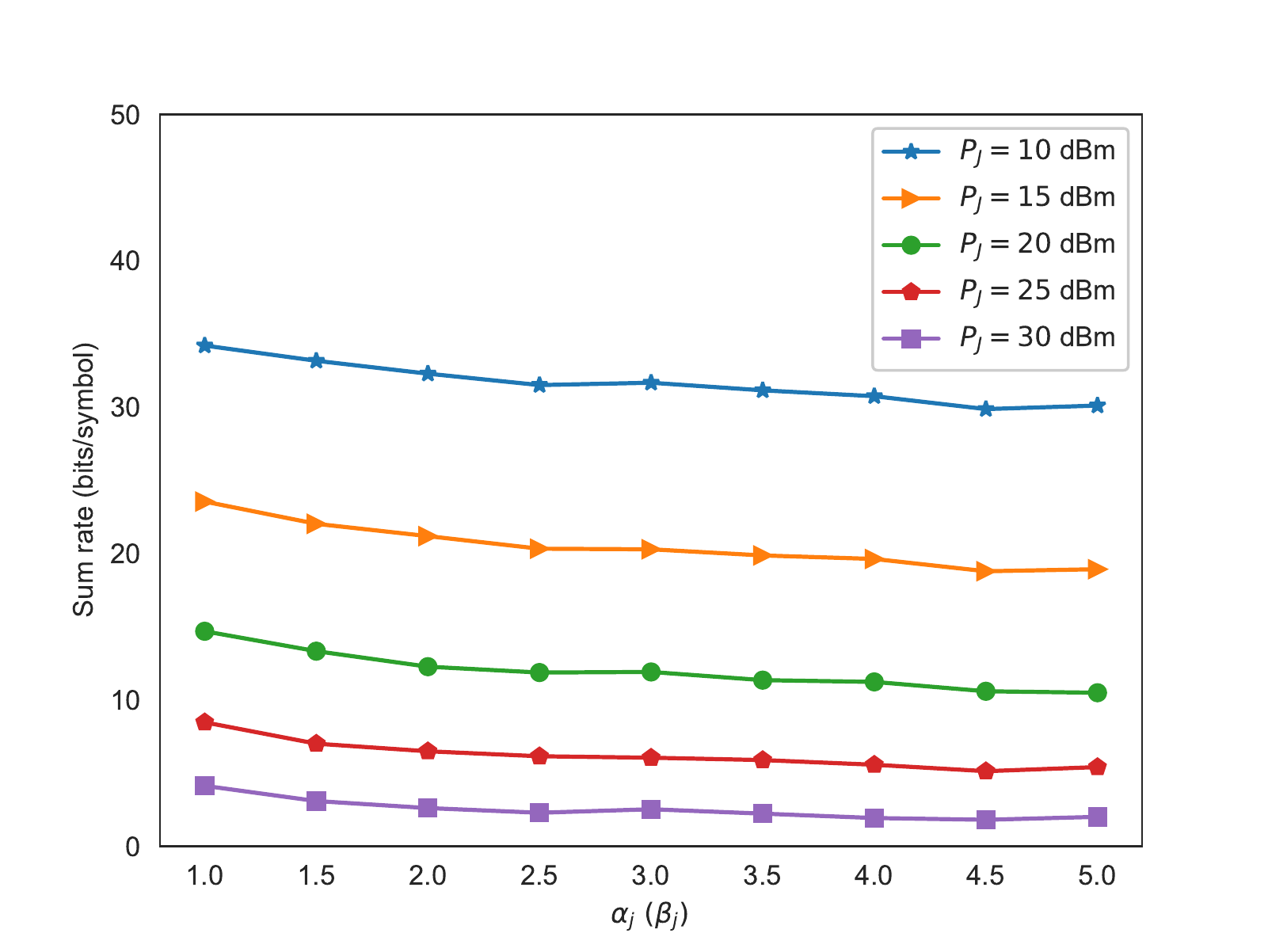}
	\caption{Sum rate vs. the shaping parameter $\alpha_j$ (or $\beta_j$). The average jamming power $P_J$ ranges from 10 dBm to 30 dBm. The jamming powers are equal across the subchannels when $\alpha_j = \beta_j = 5.0$. }
	\label{fig_rate_vs_PJ_alpha}
\end{figure}
\begin{figure}[!t]
	\centering
	\includegraphics[width=1\columnwidth]{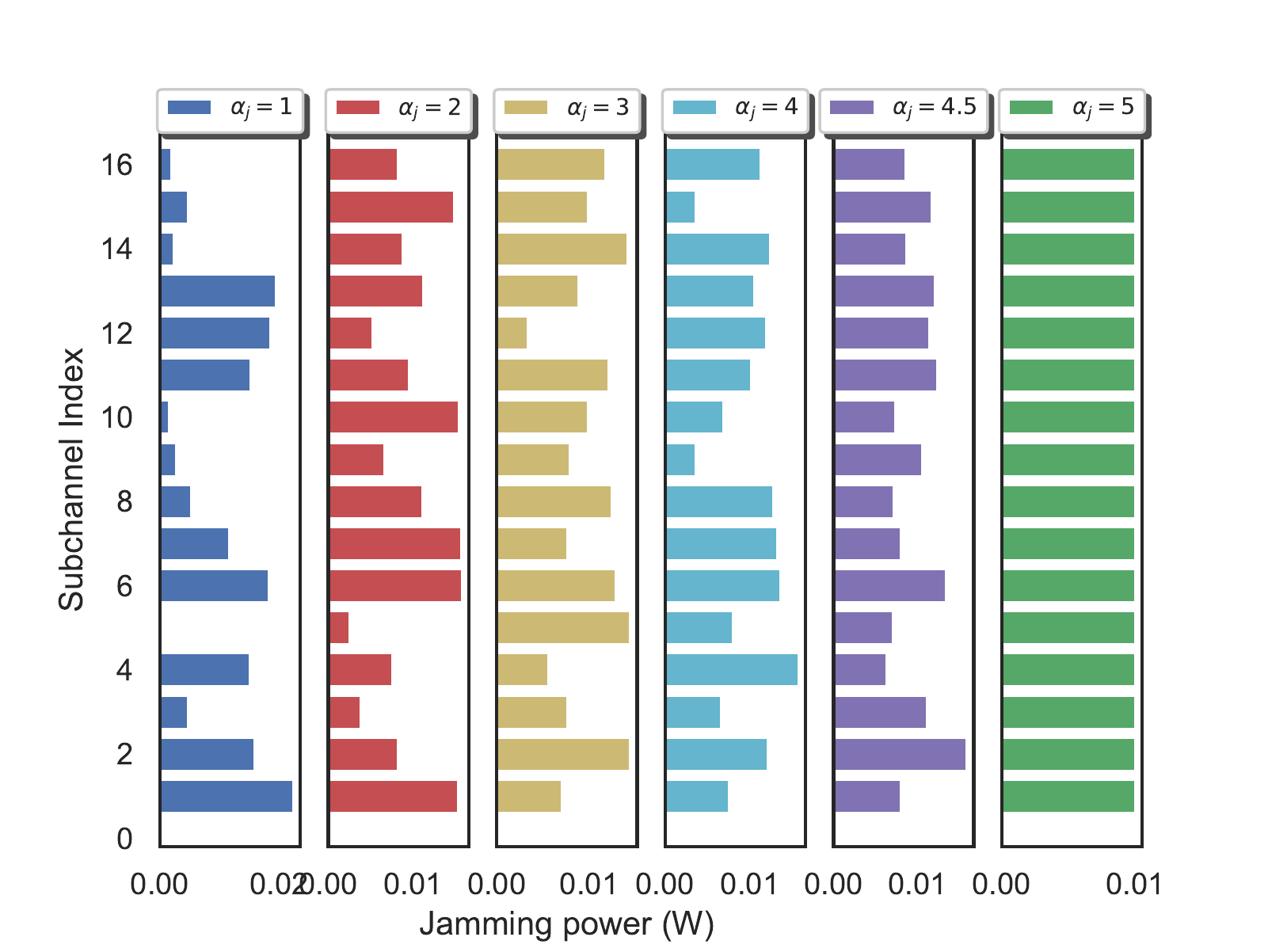}
	\caption{Examples of the unequal allocation of the jamming power $P_J$ in the subchannels under different values of $\alpha_j$ (and $\beta_j$).}
	\label{fig_PJ}
\end{figure}

To quantify the impact of unequal allocation of the jamming power on the sum-rate of the users, we project the jamming power of each subchannel using the Beta distribution, i.e., $f(x, \alpha_j,\beta_j) = x^{\alpha_j-1}(1-x)^{\beta_j-1}/B(\alpha_j,\beta_j)$, 
where $B(\alpha_j,\beta_j)$ is the Beta function with $\alpha_j = \beta_j$ being the shape parameters related to the variance of generated data. 
The larger $\alpha_j $ and $ \beta_j$ are, the most consistent the jamming power is across different subchannels. When $\alpha_j = \beta_j = 5.0$, the jamming powers are equal across the subchannels.
Fig.~\ref{fig_rate_vs_PJ_alpha} plots the sum rate against the shaping parameters $\alpha_j$ (or $\beta_j$) under different settings of the average jamming power $P_J$.
We see that the sum rate declines with the increase of $\alpha_j$ (and $\beta_j$), since the difference of the jamming power among the subchannels decreases; see Fig.~\ref{fig_PJ}.
The reason is that the unbalanced jamming powers allow the BS to avoid severely jammed subchannels and efficiently utilize those less jammed.

Fig.~\ref{fig_rate_vs_user} plots the sum rate with the increasing number of users $M$, where $N=40$, 60, and 80.
We also plot the case where the RIS is randomly configured and the case without the RIS for comparison.
It is observed that the sum rate grows with $M$ in all three cases, and the proposed PSD-TD3 outperforms the other two cases.  
The gain of the meticulously configured RIS is confirmed by showing the gain of the proposed PSD-TD3 over the case with the RIS randomly configured. 

\begin{figure}[t]
	\centering
	\includegraphics[width=0.8\columnwidth]{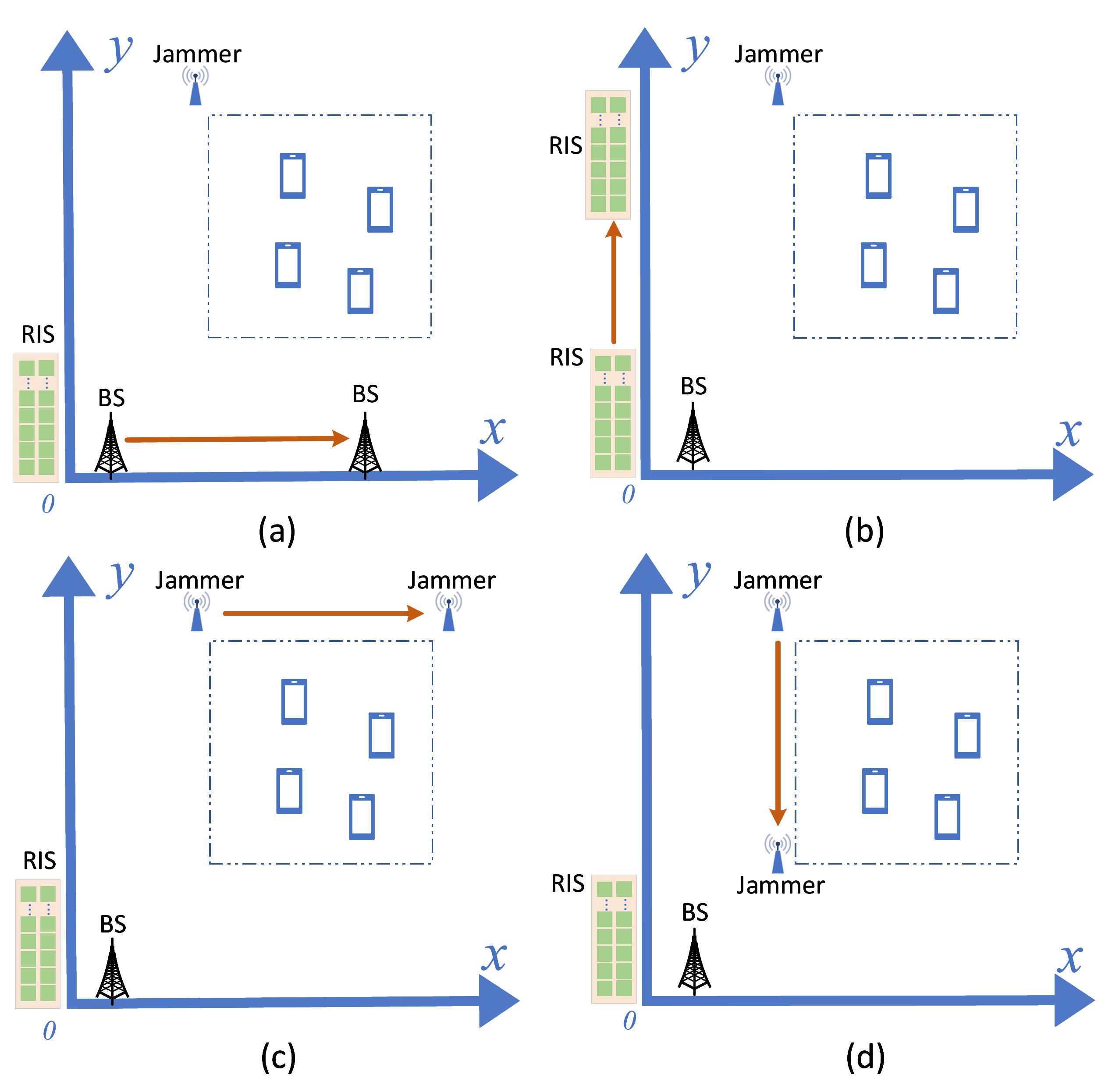}
	\caption{The bird view of the simulated system, where we assess the influence of the network deployment by moving the BS and RIS along the $x$- and $y$-axes in Figs.~\ref{fig-illustration-move}(a) and~\ref{fig-illustration-move}(b), respectively, and moving the jammer in the directions of the $x$- and $y$-axes in Figs.~\ref{fig-illustration-move}(c) and \ref{fig-illustration-move}(d).}
	\label{fig-illustration-move}
\end{figure}
\begin{figure}[t]
	\centering
	\includegraphics[width=1\columnwidth]{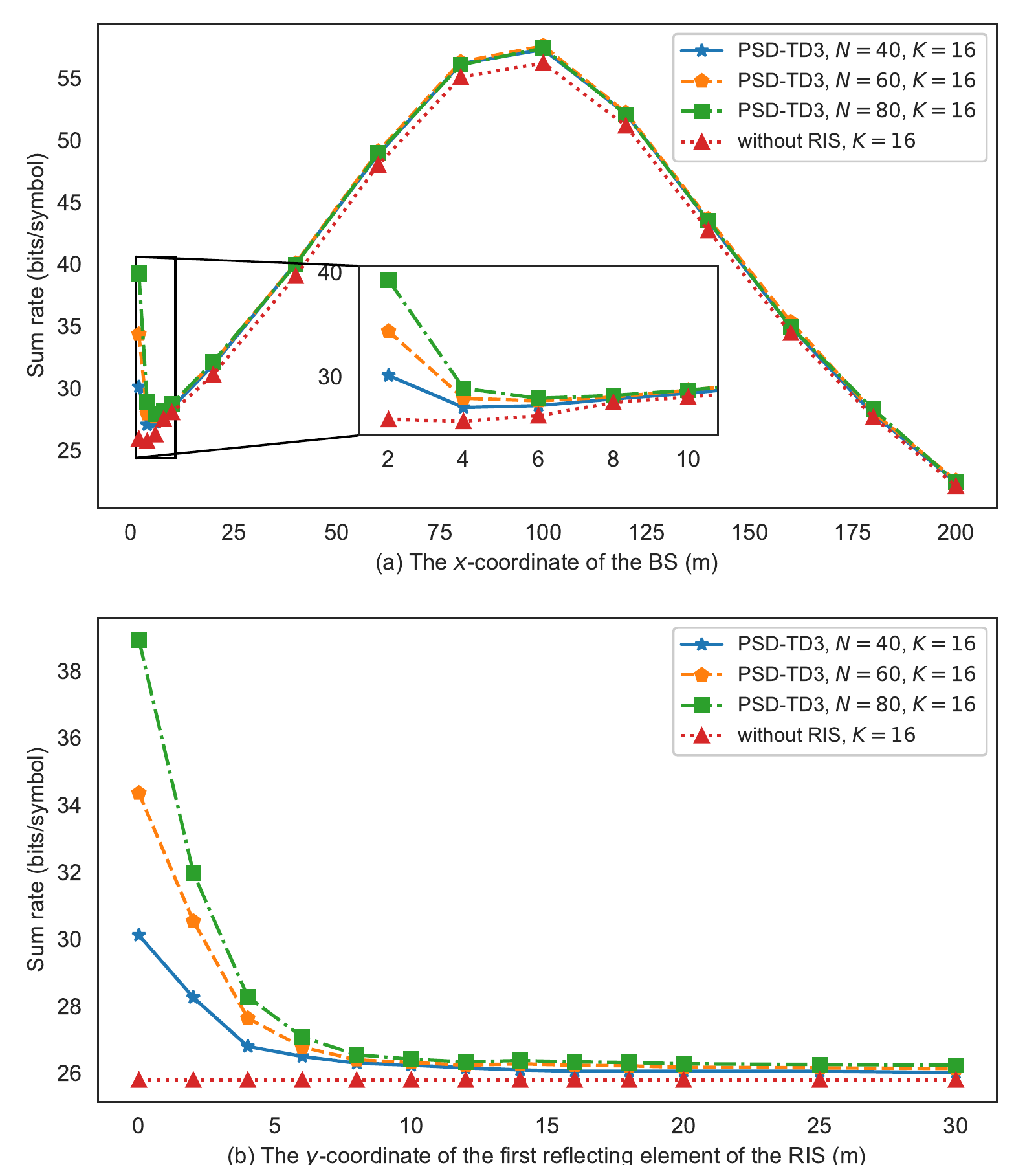}
	\caption{Sum rate vs. the horizontal and vertical distances between the BS and RIS, where we move the BS and RIS along the $x$- and $y$-axes, respectively; see Figs.~\ref{fig-illustration-move}(a) and~\ref{fig-illustration-move}(b).}
	\label{fig_rate_vs_dis}
\end{figure}
\begin{figure}[t]
	\centering
	\includegraphics[width=1\columnwidth]{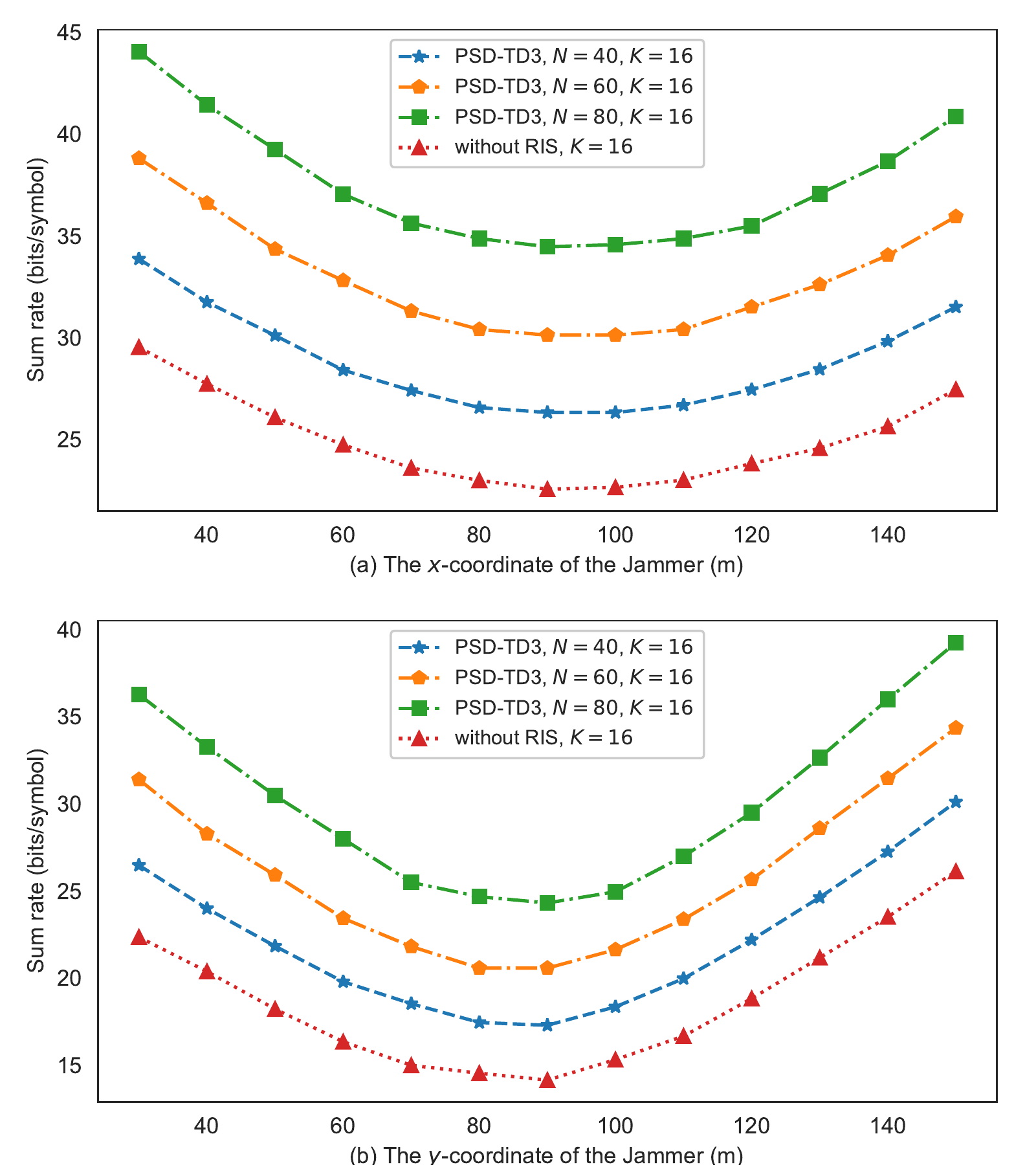}
	\caption{Sum rate vs. the vertical distance from Jammer to RIS, where we move the jammer away from the BS and RIS in the directions of the $x$- and $y$-axes; see Figs.~\ref{fig-illustration-move}(c) and \ref{fig-illustration-move}(d).}
	\label{fig_rate_vs_jammer}
\end{figure}

We proceed to assess the influence of the network deployment on the sum rate of the proposed PSD-TD3, by separately varying the positions of the BS, the RIS, and the jammer, 
as shown in Fig.~\ref{fig-illustration-move}. 
We first move the BS along the $x$-axis; see Fig.~\ref{fig-illustration-move}(a). 
Then, we move the RIS along the $y$-axis; see Fig.~\ref{fig-illustration-move}(b).
We also move the jammer along the directions parallel to the $x$- and $y$-axes; see Figs.~\ref{fig-illustration-move}(c) and~\ref{fig-illustration-move}(d).  
The results of these four cases are provided in Figs.~\ref{fig_rate_vs_dis} and \ref{fig_rate_vs_jammer}.

Fig.~\ref{fig_rate_vs_dis}(a) reveals that the sum rate of the proposed PSD-TD3 first declines quickly, then rises to its peak, and finally drops, with the increasing horizontal distance from the BS to the RIS. 
This is because less signals are reflected from the RIS, and consequently the sum rate drops rapidly as the distance starts to increase. 
By further moving the BS along the $x$-axis, the BS gets increasingly close to the users. The powers that the users receive directly from the BS increase, hence improving the sum rate. 
When the BS is moved away from the users, the received powers at the users decrease and so does the sum rate.
We also see that when the BS is in close proximity to the RIS (e.g., $D_0 \leq 5$ m), the larger number of reflecting elements at the RIS induces a higher sum rate. Nonetheless, the gain pertaining to the RIS declines when the BS is moved farther from the RIS.
Fig.~\ref{fig_rate_vs_dis}(b) shows that the sum rate declines when the RIS is moved farther from the BS along the $y$-axis (and the RIS remains far from the users). This is because the contribution of the RIS to the sum rate is increasingly negligible when the RIS is moved farther from the BS, and finally overshadowed by the contribution of the direct paths from the BS to users. 

Figs.~\ref{fig_rate_vs_jammer}(a) and \ref{fig_rate_vs_jammer}(b) show that the sum rate of the proposed PSD-TD3 first declines and then grows with the increasing vertical distances from the jammer to the RIS and the BS, respectively.
As the jammer is moved along the directions parallel to the $x$- and $y$-axes, it gets closer to the users. The received SINR at the users degrades, and hence first decreases the sum rate. 
By further moving the jammer away from the users, the jamming signal strength reduces and the sum rate increases.
We also see that the RIS-assisted system has a more powerful anti-jamming capability than the system without the RIS.
Moreover, the anti-jamming capability becomes stronger, as the number of reflecting elements increases at the RIS.

\begin{figure}[t]
	\centering
	\includegraphics[width=1\columnwidth]{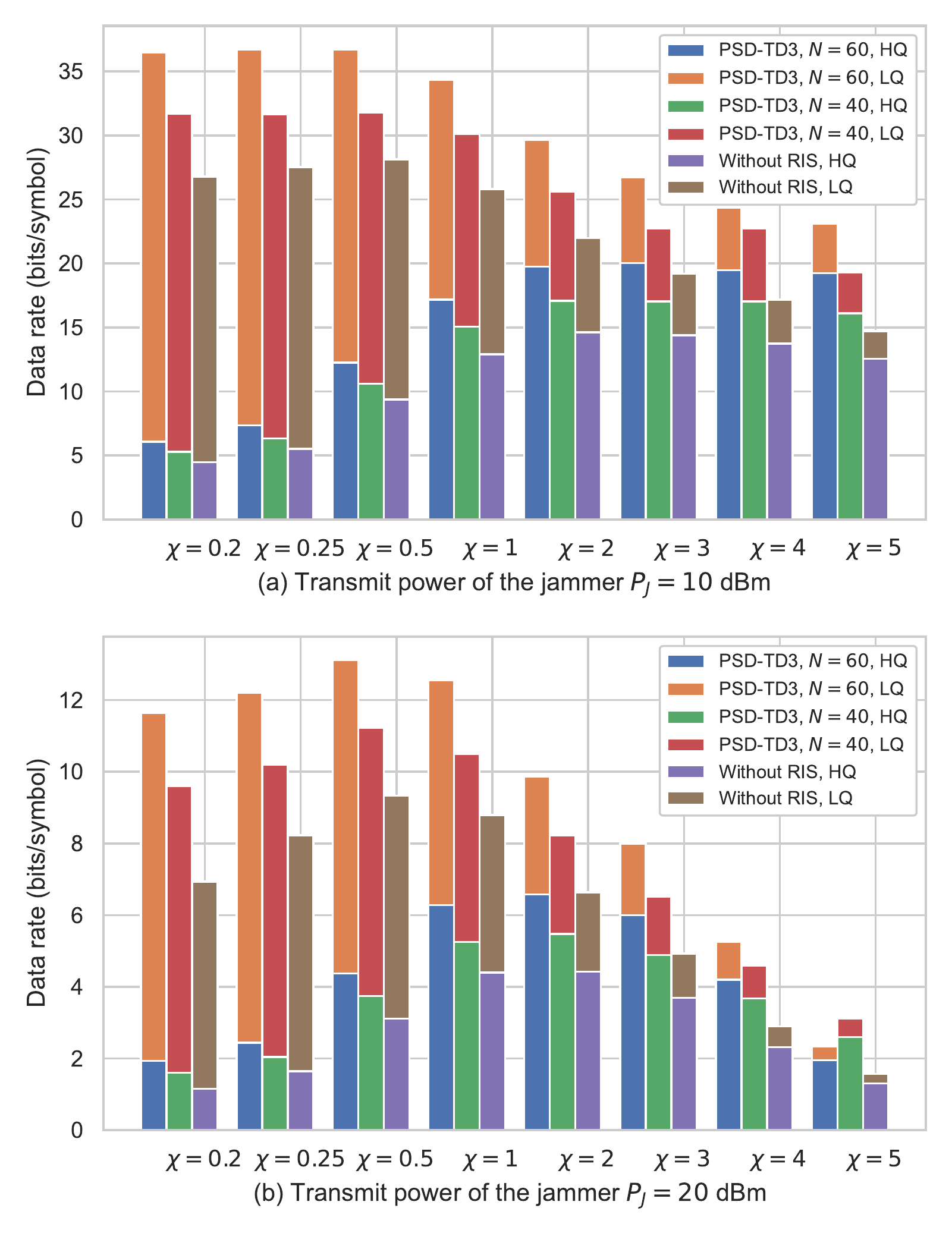}
	\caption{Sum rate vs. the ratio of the HQ and LQ data streams, where $P_{\max} = 30$~dBm. (a) The jamming power is 10~dBm. (b) The jamming power is 20~dBm.}
	\label{fig_rate_vs_chi}
\end{figure}
Finally, we assess the influence of the ratio of the high- and LQ data streams, $\chi$, on the proposed PSD-TD3 under the jammer power $P_J =10$ and 20 dBm, as shown in Fig.~\ref{fig_rate_vs_chi}.
We notice that the proposed PSD-TD3 achieves greater HQ data rates, LQ data rates, and sum rates than the case without the RIS. 
With the growth of $\chi$, the HQ data rates first grow and then decline, while the LQ data rates decrease under the proposed PSD-TD3. This is because more HQ data streams need to be delivered under a larger value of $\chi$. 
To satisfy the BER requirement (i.e., $10^{-6}$ here) of these HQ data streams, more transmit powers and channels are needed, resulting in the smaller LQ data rates and sum rates. 
On the other hand, it is increasingly difficult to satisfy the BER requirement when $\chi > 2$, owing to the unbalanced HQ and LQ data streams, and hence decreasing the HQ data rates, especially in the presence of strong jamming signals, as seen in Fig.~\ref{fig_rate_vs_chi}(b).

\section{Conclusion}\label{sec-con}
This paper proposed the new PSD-TD3 algorithm to jointly optimize user selection, channel allocation, modulation-coding adaptation, and RIS configuration for an RIS-assisted downlink multiuser OFDMA system under a jamming attack. 
A TD3 model was designed to learn the RIS configuration. The PSD was employed to optimize the user selection, channel allocation and modulation-coding adaptation. 
Both were based on the readily measurable received data rates of the users. 
Consequently, the algorithm learns to maximize the sum rate of the system through changes in the received data rates of the users, and eliminates the need of CSI. 
As validated by extensive simulations, the proposed anti-jamming PSD-TD3 framework significantly outperforms its non-learning alternatives in terms of sum rate.
The new framework with 40, 60, or 80 reflecting elements at the RIS provides 16.50\%, 32.91\%, or 51.86\% higher sum rates than the system without the RIS.

\ifCLASSOPTIONcaptionsoff
\newpage
\fi

\bibliographystyle{IEEEtran}
\bibliography{IRS_DRL_Ref_b}

\end{document}